 \documentclass[twocolumn,trackchanges]{aastex631}

\usepackage{amsmath}
\usepackage{booktabs}
\usepackage{longtable}
\usepackage{float}
\usepackage{subfigure}

\newcommand{\kms}{\mathrm{km\,s^{-1}}}

\begin{document}

\title{The Unexplored Dusty Nova LMCN 2009-05a in the Large Magellanic Cloud}

\correspondingauthor{A. Raj}
\email{ashishpink@gmail.com}

\author{Mohit Singh Bisht}
\affiliation{Indian Centre for Space Physics, 466 Barakhola, Netai Nagar, Kolkata 700099, West Bengal, India}

\author{A. Raj}
\affiliation{Indian Centre for Space Physics, 466 Barakhola, Netai Nagar, Kolkata 700099, West Bengal, India}
\affiliation{Uttar Pradesh State Institute of Forensic Science (UPSIFS), Aurawan, P.O. Banthra, Lucknow 226401 (U.P), India}


\author{F. M. Walter}
\affiliation{Department of Physics and Astronomy, Stony Brook University, Stony Brook, NY 11794-3800, USA}

\author{Anatoly S. Miroshnichenko}
\affiliation{Department of Physics and Astronomy, University of North Carolina, Greensboro, NC 27402, USA}

\author{D. Bisht}
\affiliation{Indian Centre for Space Physics, 466 Barakhola, Netai Nagar, Kolkata 700099, West Bengal, India}

\author{K. Belwal}
\affiliation{Indian Centre for Space Physics, 466 Barakhola, Netai Nagar, Kolkata 700099, West Bengal, India}

\author{Shraddha Biswas}
\affiliation{Indian Centre for Space Physics, 466 Barakhola, Netai Nagar, Kolkata 700099, West Bengal, India}






\begin{abstract}

We present a detailed spectrophotometric study of nova LMCN 2009-05a, located in the Large Magellanic Cloud (LMC). Photometric observations reveal a dust dip in the optical light curve, classifying it as a D-class nova. Light curve analysis yields $t_2$ and $t_3$ decline times of approximately 46 and 80 days, respectively, placing the nova in the category of moderately fast novae. Spectroscopic observations cover multiple phases, including pre-maximum, early decline, and nebular. The spectra are initially dominated by hydrogen Balmer and Fe II lines with P-Cygni profiles, which later transition into pure emission.  During
the optical minimum, a discrete absorption feature was observed in the
H$\alpha$ and [O I] line profiles. The physical and chemical properties during the early decline and nebular phases were analyzed using the photoionization code CLOUDY. Dust temperature, mass, and grain size were estimated through spectral energy distribution (SED) fitting to the WISE data. On day 395 post-outburst, we estimate the dust temperature to be approximately 700 K. Additionally, we examined the correlation between dust condensation time ($t_{\text{cond}}$) and $t_2$ for LMC novae, finding a trend consistent with previous studies of Galactic novae.

\end{abstract}



\section{Introduction} \label{sec:intro}

Classical novae (CNe) are thermonuclear eruptions that occur on the surface of white dwarfs in binary systems. A white dwarf in a binary system accretes matter from its non-degenerate companion star through the inner Lagrangian point, forming an accretion disk before the material is transferred to the surface of the white dwarf. 
Nuclear fusion begins once a sufficient amount of material has accumulated in the surface layers of the white dwarf. Due to its degenerate nature, this triggers a thermonuclear runaway (TNR), leading to the violent ejection of approximately 10$^{-7}$ to 10$^{-4}$ solar masses (M$_\odot$) of accreted material into the interstellar medium (ISM) at velocities ranging from a few hundred to several thousand kilometers per second \cite[][and references therein]{BodeEvansBook2008, Jose2020, Starrfield2020,chomiuk2021new}.

Novae are also identified as unique laboratories for exploring dust formation during outbursts, typically occurring within a few weeks to a few months after the eruption \citep{williams2013rapid}. Dust formation in novae is typically inferred from a dip in the optical light curve, caused by the obscuration of optical photosphere by newly formed dust, and accompanied by a simultaneous rise in infrared (IR) emission. 
Dust-dip events are associated with optically thick dust shells, where dust forms along the line of sight and significantly attenuates the optical light. In contrast, novae with an IR excess, referring to a significant increases in IR flux at epochs consistent with dust formation but without a corresponding decrease in optical brightness, are linked to optically thin dust shells, where dust does not obscure the bulk of the optical pseudo-photosphere \citep{gehrz1988infrared}.
Additional evidence for dust formation comes from asymmetries in emission line profiles. For example, \citet{shore2018spectroscopic} reported red wing suppression in the line profiles of the dusty nova V5668 Sgr and V339 Del during the onset of dust formation, shortly before the pronounced minimum observed in the optical light curve.

In the Milky Way, dust formation is frequently observed in nova eruptions \citep{evans2008dust,gehrz2008infrared}. Out of the 402 Galactic novae compiled up to 2021 by \citet{schaefer2022comprehensive}, approximately 13–14$\%$ have been reported dust formation. A similar dust-forming fraction ($\sim$14$\%$) is seen in the more recent sample of 177 novae compiled from 2008 to 2024\footnote{\url{https://asd.gsfc.nasa.gov/Koji.Mukai/novae/novae.html}}. In contrast, among the 56 recorded novae in the LMC \citep{2010AN....331..187P}\footnote{\url{https://www.mpe.mpg.de/~m31novae/opt/lmc/LMC_table.html}}, dust formation has been reported in approximately seven novae (13$\%$; \citealt{lmc1988,mroz2016ogle,2019arXiv190309232A,2025arXiv250104098C}). However, these fractions represent lower limits, as follow-up observations during the dust formation phase are often missed.
The lower number of novae in the LMC likely reflects the LMC's lower nova rate, estimated at $\sim 2.4 \pm 0.8$ yr$^{-1}$ \citep{mroz2016ogle}, which is expected given its lower mass about 10–20\%  that of the Milky Way \citep{2023Galax..11...59V}.
However, LMC novae remain particularly valuable for study due to their well-constrained distances ($d \approx 50.0 \pm 2.0$ kpc; \cite{pietrzynski2013eclipsing}) and minimal Galactic extinction along the line of sight.

Nova LMCN 2009-05a / LMC 2009 b was first reported on  2009 May 4.994 UT, by \cite{2009IAUC.9042....1L}, with a magnitude of 12.1 on two photographic frames at the position given as R.A. = 5h31m28s $\pm$ 3s, Decl. = -67$^{\circ}$05$^\prime$ 38$^{\prime\prime}$  $\pm$ 8$^{\prime\prime}$. No object was visible at this location in their
frames taken on 2009 April 23.01 UT (limiting magnitude 14). CCD images taken on May 7.716 UT by L. A. G. Monard reported a confirmation of the presumed nova at an unfiltered mag 12.5, with the position R.A. = 5h31m26.42s, Decl. = -67$^{\circ}$05$^{\prime}$ 39.4$^{\prime\prime}$ . A search for the pre-eruption object in the Digitized Sky Survey (DSS) (limiting red magnitude $\sim$ 20) from 1991 and the U.K. Schmidt image of the field (limiting magnitude $\sim$19) from 1996 did not reveal any object at the position of LMCN 2009-05a.
In the later stages, the nova was caught in the Optical Gravitational Lensing Experiment (OGLE) survey during its decline phase, between 2010 March 5.15 and 2014 March 6.12 UT \citep{mroz2016ogle}.

In this paper, we present a detailed study of the nova LMCN 2009-05a. The paper is organized as follows: Section \ref{observations} describes the observations. Section \ref{analysis} discusses the optical light curve evolution, including the dust dip, and presents the spectral evolution during the pre-maximum rise, early decline, and nebular phases. This section also includes the estimation of key physical and chemical parameters through photoionization modeling, as well as dust properties such as mass and grain size using Wide field Infrared Survey Explorer (WISE)
data. Section \ref{Discussion} provides a discussion on the relation between {dust condensation time ($t_{\text{cond}}$) and the time taken to decline by 2 magnitudes from maximum light ($t_2$) for both LMC and Galactic novae. A summary is presented in Section \ref{summary}.

\section{Observations}\label{observations}
All photometric data used in this study were obtained from the American Association of Variable Star Observers (AAVSO\footnote{\href{https://www.aavso.org/}{https://www.aavso.org/}}) international database and the Small and Medium Aperture Telescope System (SMARTS\footnote{\href{http://www.astro.sunysb.edu/fwalter/SMARTS/NovaAtlas/}{http://www.astro.sunysb.edu/fwalter/SMARTS/NovaAtlas/}}). Additionally, a total of 34 spectra were acquired using the SMARTS R/C spectrograph between May 12, 2009 (day 8) and January 20, 2010 (day 261), with an irregular observational cadence. Details about the SMARTS R/C grating spectrograph, data reduction methods, and observing modes can be found in \cite{wal12}. The spectroscopic observation log is presented in Table \ref{log}.

\begin{table}
\centering
\caption{Observational log for spectroscopic data obtained for LMCN 2009-05a.}
\label{log}
\resizebox{\hsize}{!}{%
\begin{tabular}{lccccc}
\hline
\hline
& \textbf{Time since}  & \textbf{Exposure}  & \textbf{Wavelength} & \textbf{Resolution} \\
\textbf{Date (UT)} & \textbf{discovery} & \textbf{time}  & \textbf{range} & \textbf{(\AA)} \\
& \textbf{(days)} & \textbf{(s)} & \textbf{(\AA)} &   \\
\hline
			2009 May 12.99 & 8.00 & 600  & 3647--5415 & 4.1 \\[0.25ex]
			2009 May 14.96 & 9.97 & 900  & 5628--6944 & 3.1  \\[0.25ex]
			2009 May 26.95 & 21.95 & 300 & 3649--5415 & 4.1  \\[0.25ex]
			2009 May 31.95 & 26.95 & 800  & 3649--5415 & 4.1 \\[0.25ex]
			2009 July 07.44 & 63.45 & 600 & 3649--5415 & 4.1  \\[0.25ex]
			2009 July 12.42 & 68.43 & 720  & 5627--6944 & 3.1  \\[0.25ex]
			2009 July 16.42 & 72.41 & 720  & 3871--4544 & 1.6 \\[0.25ex]
			2009 July 19.41 & 75.42 & 720  & 3650--5420 & 4.1  \\[0.25ex]
			2009 July 23.41 & 79.41 & 2700   & 3400--9610  & 17.2 \\[0.25ex]
			2009 July 26.42 & 82.42 & 2700 & 5629--6947  & 3.1 \\[0.25ex]
			2009 July 28.41 & 84.41 & 600  & 5629--6947 & 3.1  \\[0.25ex]
			2009 July 31.37 & 87.37  & 600  & 5630--6949 & 3.1  \\[0.25ex]
			2009 August 03.37 & 90.37  & 600  & 5630--6949 & 3.1  \\[0.25ex]
			2009 August 06.35 & 93.35  & 600  & 5630--6949 & 3.1  \\[0.25ex]
			2009 October 09.27 & 157.27  & 600  & 3652--5420 & 4.1   \\[0.25ex]
			2009 October 15.17 & 163.17  & 600  & 5632--6951 & 3.1   \\[0.25ex]
			2009 October 18.27 & 166.27  & 600  & 3660--5430 & 4.1  \\[0.25ex]
			2009 October 20.19 & 168.20  & 600  & 5631--6944  & 3.1  \\[0.25ex]
			2009 October 23.36 & 171.36  & 600  & 3400--9610  & 17.2   \\[0.25ex]
            2009 October 25.34 & 173.34  & 600  & 6007--9483  & 6.5   \\[0.25ex]
			2009 October 26.35 & 174.36  & 600  & 6007--9483  & 6.5   \\[0.25ex]
			2009 October 31.14 & 179.15  & 600  & 5629--6947   & 3.1  \\[0.25ex]
			2009 November 01.19 & 181.20  & 600  & 5631--6950   & 3.1\\[0.25ex]
			2009 November 06.13 & 185.13  & 600  & 3653--5420  & 4.1  \\[0.25ex]
			2009 November 08.18 & 187.19  & 600  & 5631--6948  & 3.1  \\[0.25ex]
			2009 November 18.18 & 197.19  & 600  & 5629--6948  & 3.1  \\[0.25ex]
			2009 November 27.19 & 206.19  & 600  & 6241--7565  & 3.1  \\[0.25ex]
			2009 December 17.09 & 226.10  & 600  & 3649--5418  & 4.1  \\[0.25ex]
			2009 December 26.06 & 235.06  & 600  & 5632--6949  & 3.1  \\[0.25ex]
			2009 December 27.14 & 236.15  & 600  & 3400--9610  & 17.2 \\[0.25ex]
			2009 December 30.06 & 239.06  & 600  & 6233--7556 & 3.1  \\[0.25ex]
			2010 January  12.04 & 252.05  & 600  & 5624--6942 & 3.1  \\[0.25ex]
			2010 January  20.04 & 260.05  & 600  & 5995--9477 & 6.5  \\[0.25ex]
			2010 January  21.07 & 261.07  & 600  & 3647--5415 & 4.1  \\[0.25ex]
\hline
\end{tabular}}
\end{table}

\section{Analysis}\label{analysis}

\subsection{Optical light curve, dust dip, and outburst luminosity}\label{optical light curve}
The optical light curves based on the data from the AAVSO database and SMARTS are presented in Fig. \ref{lc_optical}. Nova LMCN 2009-05 was first reported on 2009 May 4.994 UT; in this paper, we considered this date as the day of the outburst (day 0). The BVRI band light curve begins within 3.5 days after outburst. The brightness increased at a slow rate for all bands and reached the peak magnitude V$_{max}$ =12.29 $\pm$ 0.01 after 15.5 days. The brightness in the BVRI bands shows a slow decline after the peak.  
A sudden decrease in the brightness in the BVRI bands was observed on day 89, indicating obscuration of optical flux by dust and suggesting that a significant amount of dust had formed in the nova ejecta. This is further supported by near-infrared (NIR) observations, which show an increase in the brightness of the NIR J, H, and K bands (see Section \ref{nir_lc} for details on the NIR data). The brightnesses in the B, V, R, and I bands were 14.61 $\pm$ 0.01, 14.62 $\pm$ 0.01, 14.09 $\pm$ 0.01, and 13.99 $\pm$ 0.01 mag on day 67, and decreased to 18.46 $\pm$ 0.11, 18.49 $\pm$ 0.15, 17.08 $\pm$ 0.05, and 16.76 $\pm$ 0.06 mag, respectively, on day 89.
There were no observations between days 67 and 89, so the onset of dust formation is taken to be around 78 $\pm$ 10 days since outburst. The minima of the optical dip were observed on day 108 in the B band at 19.04 $\pm$ 0.1 and on day 126 in the V, R, and I bands. After the minima, all bands began to recover and on day $\sim$155, they had fully recovered. After 155 days, the brightness started to slowly decrease and entered the final decline phase. On the final observation (day 3591) of the optical light curve, the nova had magnitudes of 19.80 $\pm$ 0.13, 19.05 $\pm$ 0.08, 18.65 $\pm$ 0.06, and 18.24 $\pm$ 0.08 in the B, V, R, and I bands, respectively.

The speed class of novae is usually described by t$_2$ , the time to decline by 2 mag from maximum light. From the least-squares regression fit to the V band light curve, we estimate t$_2$ to be 46 $\pm$ 3 days. Based on the $t_2$ value the nova belongs to a moderately fast nova category \citep{1957gano.book.....G}.
Due to the dust-dip in the photometric data, we estimated $t_3$ to be 80 ± 5 days using the relation $t_3$ = 2.75 ($t_2$)$^{0.88}$ by \cite{1995cvs..book.....W}. A classification system for the optical light curves
for novae on the basis of the shape of the light curve and the time to decline by 3 mag ($t_3$) from $V_{max}$ has been presented by \cite{strope2010catalog}. The shape of the optical light curve of LMCN 2009-05a presented in Fig. \ref{lc_optical} has all the characteristics of the D class of nova, which shows a dust dip in the optical light curve after the optical maximum. For the LMC, we use the distance modulus
of $\mu_0$=18.49 ± 0.05 \citep{pietrzynski2013eclipsing} and the reddening of E (B-V) =0.13 (see Section \ref{reddening and distance}), to calculate the maximum absolute magnitude of the nova to be M$_V$ = -6.65 ± 0.06. By utilizing the values of M$_V$
and t$_3$ and applying them to the equations 2 and 6 given in \cite{livio1992classical}, we obtained the mass of WD (M$_{WD}$) $\sim$ 0.77 $\pm$ 0.10M$_\odot$. This indicates that the nova LMCN 2009-05a contains low-mass CO WD (M$_{WD}$ $\leq$  1.2 M$_\odot$). 
The outburst luminosity is estimated from M$_{bol}$ = 4.8 + 2.5log (L/L$_\odot$) where the bolometric correction applied to M$_V$ is assumed to lie between -0.4 and 0.00, corresponding from the A to F spectral types \citep{gehrz1988infrared}. 
Using M$_V$ = -6.65, we estimate the outburst luminosity to be (4.64 ± 0.65)× 10$^{4}$ L$_\odot$. This means that LMCN 2009-05a was a relatively low-luminosity nova
(see, e.g., Fig. 2 in \cite{shara2017hubble}).

The evolution of V-R and V-I colors is shown in Fig. \ref{op_color}. The V-R and V-I colors gradually increased from approximately 0.41 to 0.52 and 0.58 to 0.63, respectively, from day 11 to day 67. During the dust formation phase on day 89, both V-R and V-I showed a sharp increase, reaching 1.42 and 1.74, respectively. Thereafter, the color indices gradually decreased.
The data point on day 126 showed a significantly larger uncertainty in V-R (0.49 mag) and V-I (1.09 mag). These increased errors reflect the non-photometric conditions during that observation (see insets in Figure \ref{lc_optical}).

\begin{figure*}
	\includegraphics[scale=0.55]{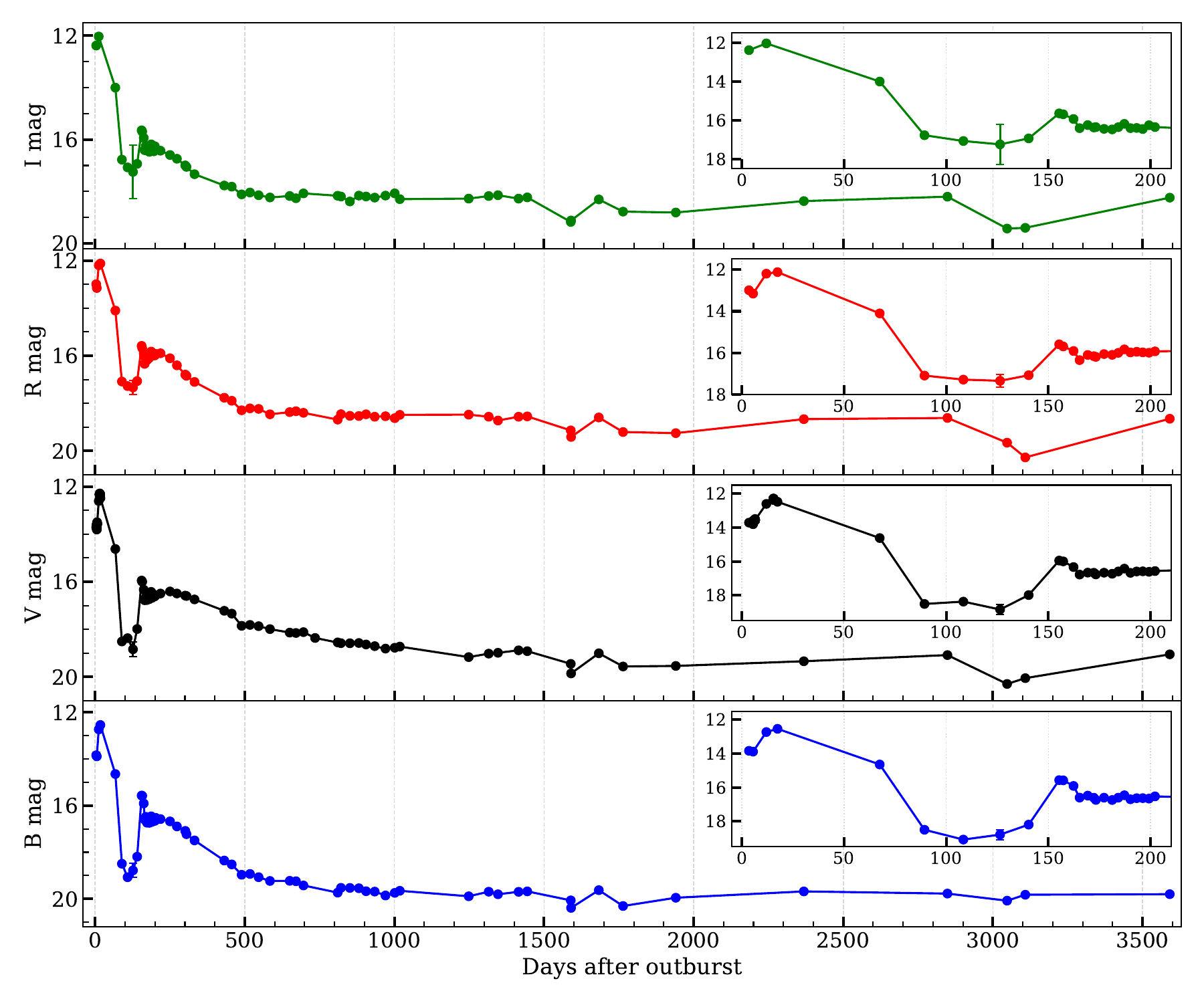}
\caption{Optical light curves of LMCN 2009-05a from day 3.5 to 3591 generated using optical data from AAVSO
and SMARTS. An inset highlights the first 200 days, showing the light curve behavior during the optical minimum, indicating obscuration of optical flux by dust (see Section \ref{optical light curve} for more details).
}
	\label{lc_optical}
\end{figure*}

\begin{figure}
	\includegraphics[scale=0.36]{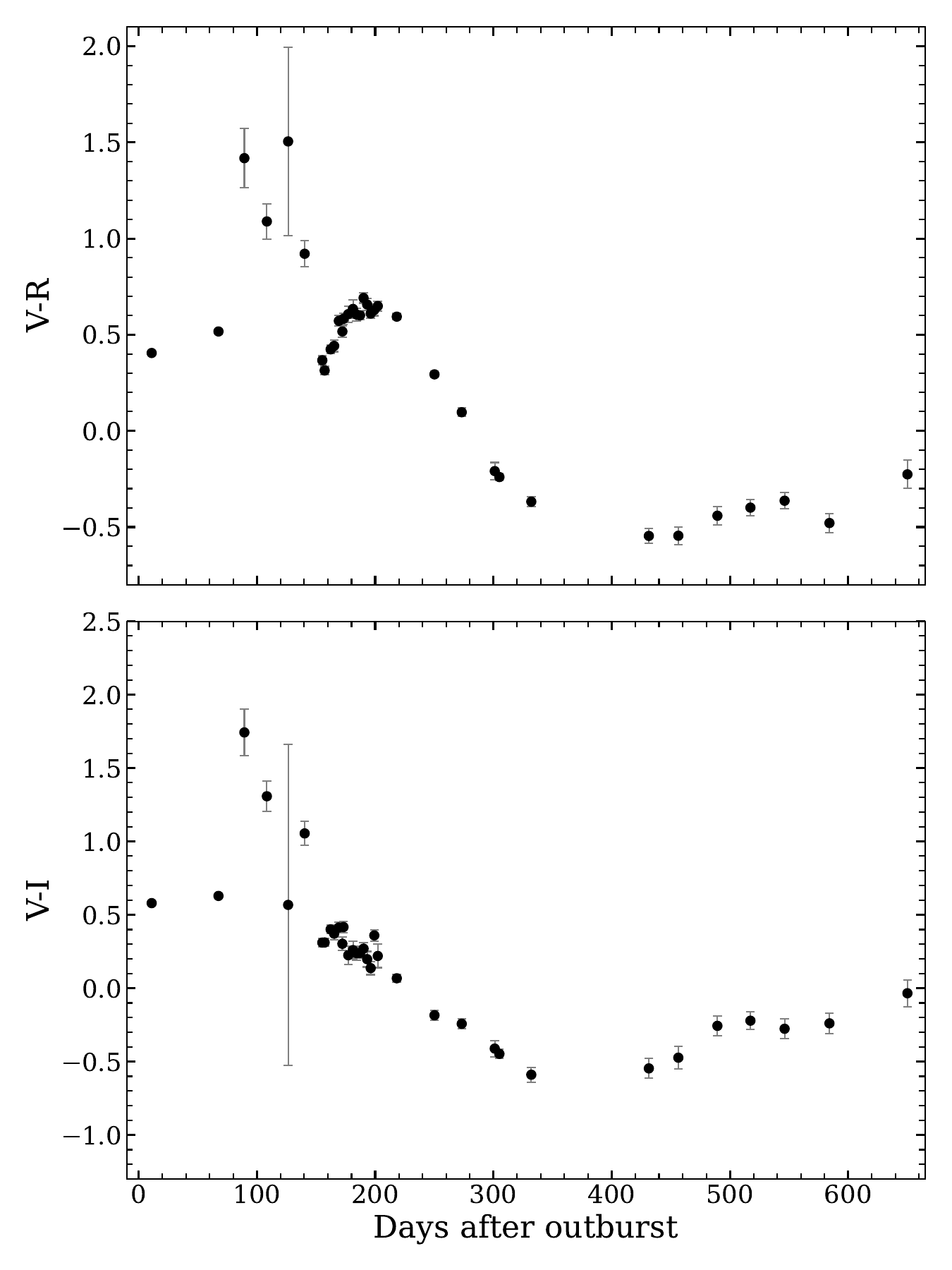}
\caption{Optical color of LMCN 2009-05a from day 11 to 650 generated using SMARTS. The large error bars near day 126 correspond to data obtained under non-photometric conditions.}
	\label{op_color}
\end{figure}

\begin{figure*}
\includegraphics[scale=0.55]{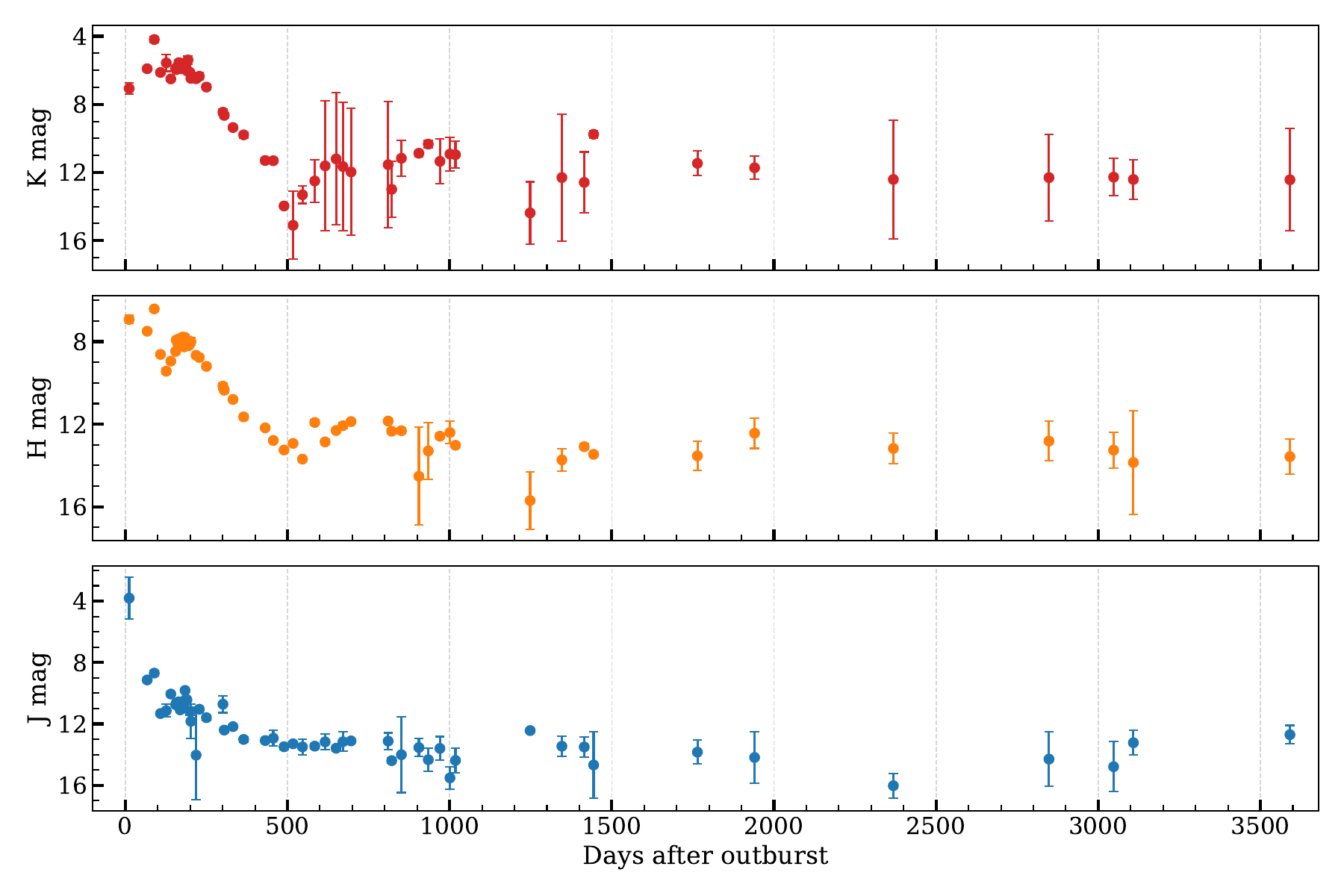}
\caption{NIR light curves of LMCN 2009-05a from day 11 to 3591 generated using SMARTS.}
	\label{lc_nir}
\end{figure*}

\subsection{NIR light curve}\label{nir_lc}

The NIR light curves are made using data from SMARTS are presented in Fig. \ref{lc_nir}. The evolution of the J-H, H-K and J-K color indices is shown in
Fig. \ref{nir_color}. These indices are clearly influenced by the presence
of dust around the system. The J-K, J-H and H-K colors
reach a maximum value of 5.6, 2.7, and 3.8 mag, respectively after 67 days. The maximum values of these colors indicate substantial dust formation within the ejecta between 67 and 155 days. After this period, the color indices decreased. 
This is because due to thermal emission from dust during the condensation phase, the longer wavelength fluxes increase, followed by a rapid decline after grain growth ceases, as the shell density decreases due to expansion \citep{gehrz2008infrared}. High J-K values are also observed for other dust-forming novae; 3.79 for V496 Sct \citep{raj2012v496}, 8.00 for V2676 Oph \citep{kawakita2017mid,raj2017v2676}, and 6.7 for V5579 Sgr \citep{raj2024dustyaftermathrapidnova}. In the final NIR light curve observation (day 3591), the nova had magnitudes of 12.70 $\pm$ 0.6 , 13.57 $\pm$ 0.8, and 12.43 $\pm$ 2.8 in the J, H, and K bands, respectively.

\begin{figure}
	\includegraphics[scale=0.45]{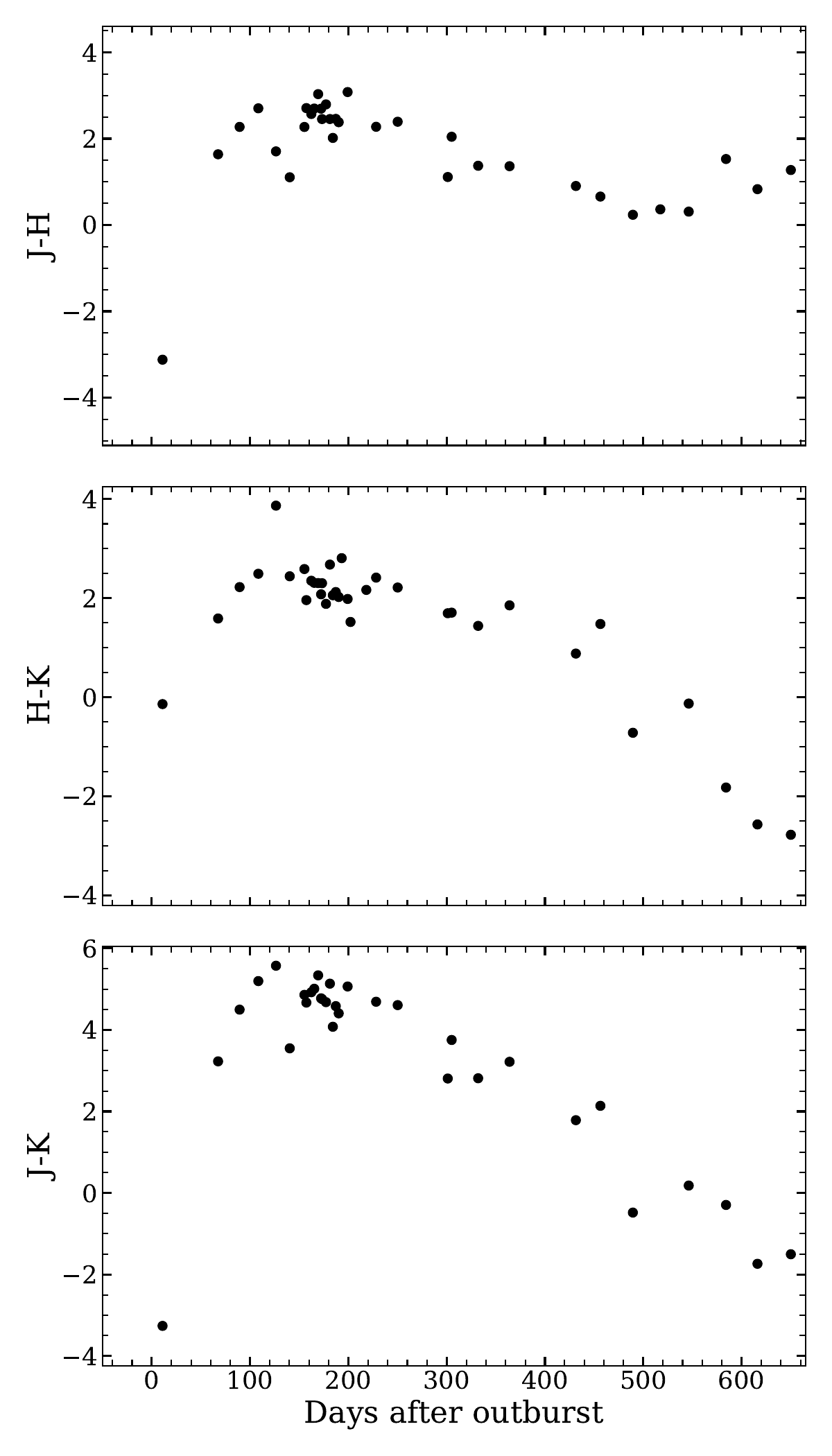}
\caption{NIR color of LMCN 2009-05a from day 11 to 650 generated using SMARTS.}
	\label{nir_color}
\end{figure}

\subsection{Reddening and distance}\label{reddening and distance}

The Galactic reddening toward nova LMCN 2009-05a is E(B-V) = 0.07, as given by \cite{schlafly2011measuring}, based on a recalibration of the Galactic reddening maps by \cite{schlegel1998maps}. This value provides a lower limit for any sufficiently distant object in the direction of nova LMCN 2009-05a. The reddening range for the LMC varies across studies. \cite{imara2007extinction} reported E(B-V)=0.12, \cite{schwarz2011swift} used E(B-V)=0.15, and \cite{2020A&ARv..28....3D} estimated E(B-V)=0.14, derived from six novae as reported by \cite{2013AJ....145..117S}. \cite{2019A&A...628A..51J} reported E(B-V) = 0.1 at approximately 6.2 arcminutes from the LMCN 2009-05a, based on OGLE IV Cepheids. We adopted an average reddening value of E(B-V) = 0.13 $\pm$ 0.02. Using this, we derived the interstellar extinction as A$_V$ = 0.44 for R$_V$ = 3.4 \citep{2023ApJ...946...43W}. 
The distance to the LMC is already well-known, d= 50$\pm$2 kpc \citep{pietrzynski2013eclipsing}. In this paper, we assume this distance for the nova.

\subsection{Line identification, general characteristics, and evolution of the optical spectra}\label{Spectral evolution}
The spectroscopic evolution of LMCN 2009-05a, covering the pre-maximum phase, the early decline phase (including the optical minimum), and the nebular phase, is presented in Figs. \ref{spectral1}, \ref{spectral2}, and \ref{spectral3}. The prominent emission features are identified based on \cite{williams2012origin}. We present 34 low-dispersion spectra spanning from 2009 May 12 to 2010 January 21 (day 8 to 261). The pre-maximum spectra taken on day 8 and 10 showed emission from Balmer lines (H$\alpha$, H$\beta$, H$\gamma$, H$\delta$), lines of Fe II multiplets (4924, 5018, 5169 \AA) along
with Ca II (H and K). All these lines exhibited deep P-Cygni profiles with an absorption component. The absorption minimum of the Balmer and Fe II lines was blue-shifted from the emission peak by $-790 \pm 40 \, \mathrm{km\,s^{-1}}$.
Other lines include Na I 5892 \AA, Na I 6154/6160 \AA, and [O I] 6300 \AA.

The red spectra taken on day 22 and 26 (6.45, 11.45 days post optical maxima) showed same emission lines as previous epochs with a strong P-Cygni absorption component. The absorption minimum of the Balmer and Fe II lines was blue-shifted from the emission peak by $-895 \pm 40 \, \mathrm{km\,s^{-1}}$. On days 63, 68, 72 and 75, the P-Cygni profiles are still clearly seen, indicating a slowly moving ejecta. On day 63 and 68 the Balmer lines showed strongest P-Cygni absorption component with absorption at $-1240 \pm 60$ $\kms$ from emission peak. On day 68 [O I] 6300 and 6364 \AA{} are both seen. The spectrum taken on day 79 showed clear emission. The lines include emission from Balmer lines, Fe II multiplets (27, 38, 42, 48, and 49), and O I 7773, 8446 \AA. The forbidden lines of oxygen [O I] 5577, 6300, 6364 \AA{} and [N II]
5755 \AA{} indicate the inhomogeneous density structure of the ejecta, with a wide range of densities \citep{williams2013novae}. The FWHM velocities of H$\alpha$, [O I] , [N II] were measured to be 1050 $\kms$, 720 $\kms$, 1100 $\kms$, respectively.

The photometric data indicates the onset of a dust dip between days 67 and 89, with the B-band magnitude decreasing by approximately 4.43 mags over $\sim$ 40 days, ultimately reaching a minimum brightness of B = 19.04 $\pm$ 0.1 on day 108. Spectra obtained on days 82, 84, 87, 90, and 93 revealed a noticeable discrete absorption feature in the H$\alpha$ and [O I] 6300, 6364 \AA{} profiles, centered at approximately +300 $\kms$ in velocity space, corrected for the LMC velocity (278 $\kms$; \cite{1987A&A...171...33R}; see Fig.~\ref{radial}), while the FWZI remained unchanged during this period.
Similar profiles were also observed in the case of V5668 Sgr \citep{gehrz2018temporal}, V339 Del \citep{shore2018spectroscopic}.  
Such a behavior is consistent with dust formation scenarios within the nova ejecta, where asymmetry in the line profiles indicates the presence of dust that preferentially obscures the receding (redshifted) material.
The persistence of blueward emission suggests that the approaching material was less affected by dust extinction \citep{shore2018spectroscopic}. However, in the subsequent observing season, a reversal feature reappeared at a slightly lower velocity (see Fig.~\ref{h alpha}). 
Such asymmetrical profile reversals are commonly seen during the nebular phase \citep{chomiuk2021new} and are likely due to emission from a geometrically thin shell, consistent with the interpretation of aspherical or bipolar ejecta structures \citep{hutchings1972non,2012ApJ...755...37H,ribeiro2013optical,2018ApJ...853...27M}.

The dust was no longer optically thick in the optical by around day 155, as indicated by the photometric points, with V-band magnitudes of 17.98 $\pm$ 0.06 on day 140, 15.95 $\pm$ 0.02 on day 155, and 16.32 $\pm$ 0.02 on day 162. The spectra taken in October 2009 on day 157, 163, 166, and 168 showed nebular lines [O III] 4363, 4959, 5007 \AA{} followed by lines of [N II] 5755, and blend at 4640 \AA( N III 4638 + [Fe III] 4658+ He I 4686 \AA). The full range spectra taken on day 171 showed the strongest emission lines in the spectra arise from H$\alpha$ followed by [O III] 5007, blend 4640, H $\beta$, [O III] 4363, 4959, [N II] 5755, and [O II] 7320 \AA.
The spectra taken in October 2009 (days 157 to 179) showed lines of helium and nitrogen such as; N II 5679, He I 5876, He I 6678, He I 7065 and weak C II 4267 \AA. The above lines showing that the nova has gone through the later He/N phase \citep{2024MNRAS.527.9303A} after which it entered into the nebular phase. The spectrum taken on day 185 showed that the line fluxes of the nebular lines ([O III] 4363, 4959, 5007 \AA{}) had increased. 
In December 2009, the full spectrum taken on day 236 showed the strongest lines: H$\alpha$, followed by [O III] 5007, 4363, H$\beta$, [O III] 4959, and [N II] 5755 \AA. 
In the final spectra taken at 260 and 261 days, nebular lines dominated the spectra. The 4640 blend was resolved, with both N III 4638 and He I 4686 \AA{} clearly visible. The [O III] 4363 \AA{} line prominently dominated the blend with the H$\gamma$ line.

The evolution of the velocity profile of H$\alpha$ and [O I] 6300 \AA{} is presented in Fig. \ref{h alpha}.

\begin{figure*}
\centering
	\includegraphics[scale=0.54]{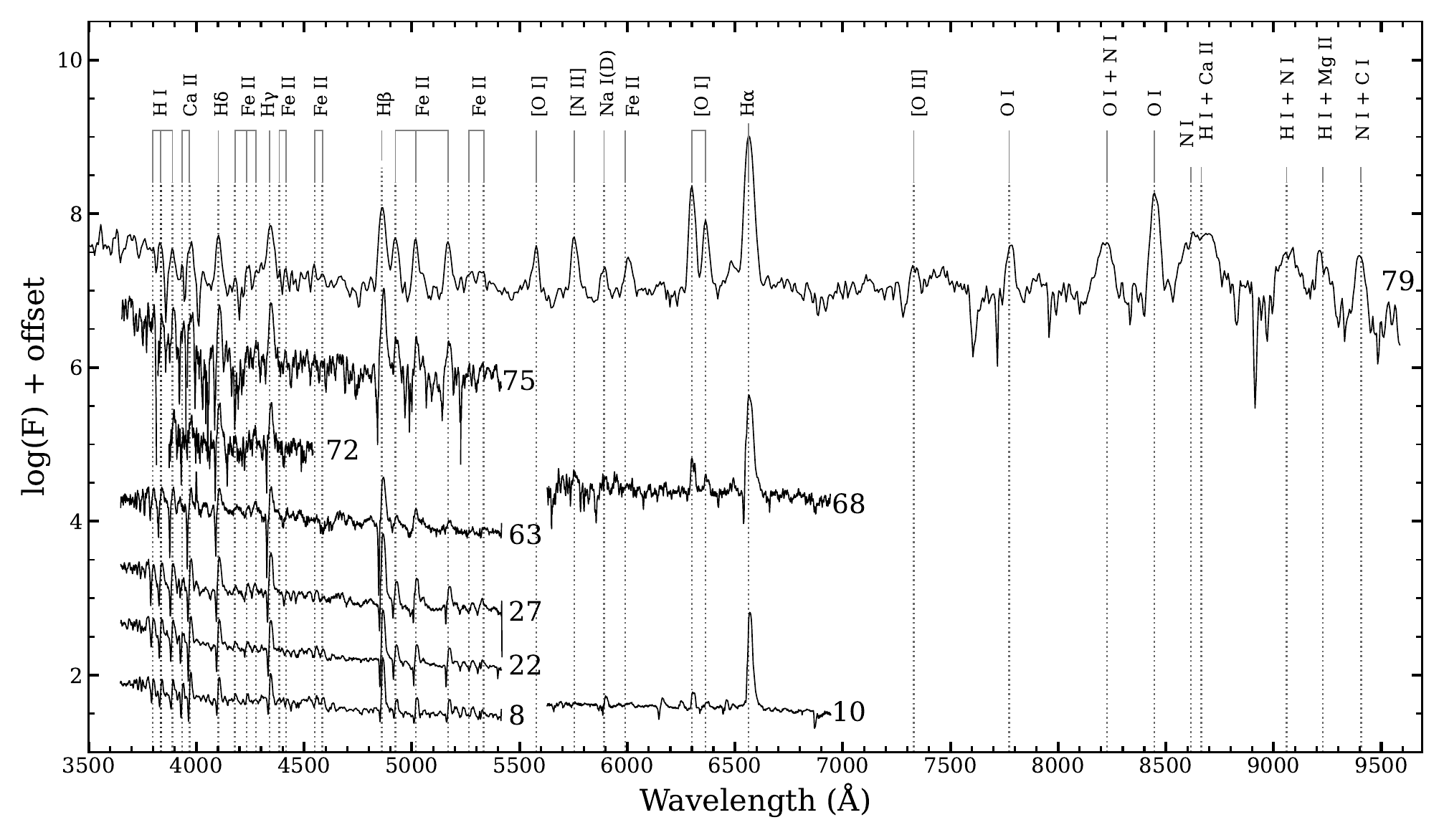}
\caption{Early spectroscopic evolution of Nova LMCN 2009-05a, obtained from day 8 (2009 May 12) to day 79 (2009 July 23). The spectra prominently feature Fe II multiplets and hydrogen Balmer lines. Identified lines are marked, and the time since discovery (in days) is labeled beside each spectrum. }
	\label{spectral1}
\end{figure*}

\begin{figure*}
\centering
    \includegraphics[scale=0.54]{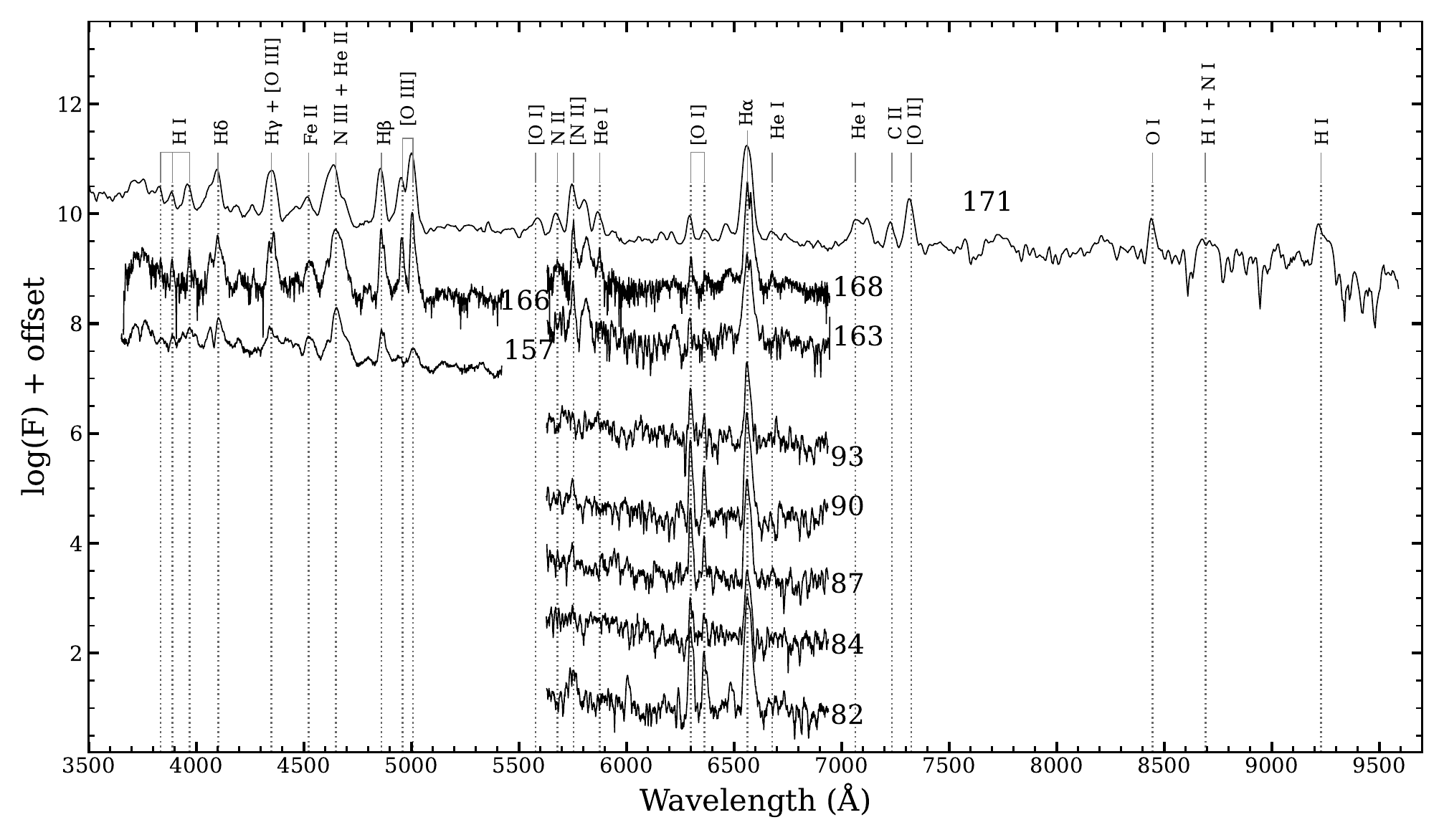}
\caption{Spectroscopic evolution of Nova LMCN 2009-05a during the dust formation phase, from day 82 (2009 July 26) to day 171 (2009 October 23). Identified lines are marked, and the time since discovery (in days) is labeled beside each spectrum. }
	\label{spectral2}
\end{figure*}

\begin{figure*}
\centering
	\includegraphics[scale=0.54]{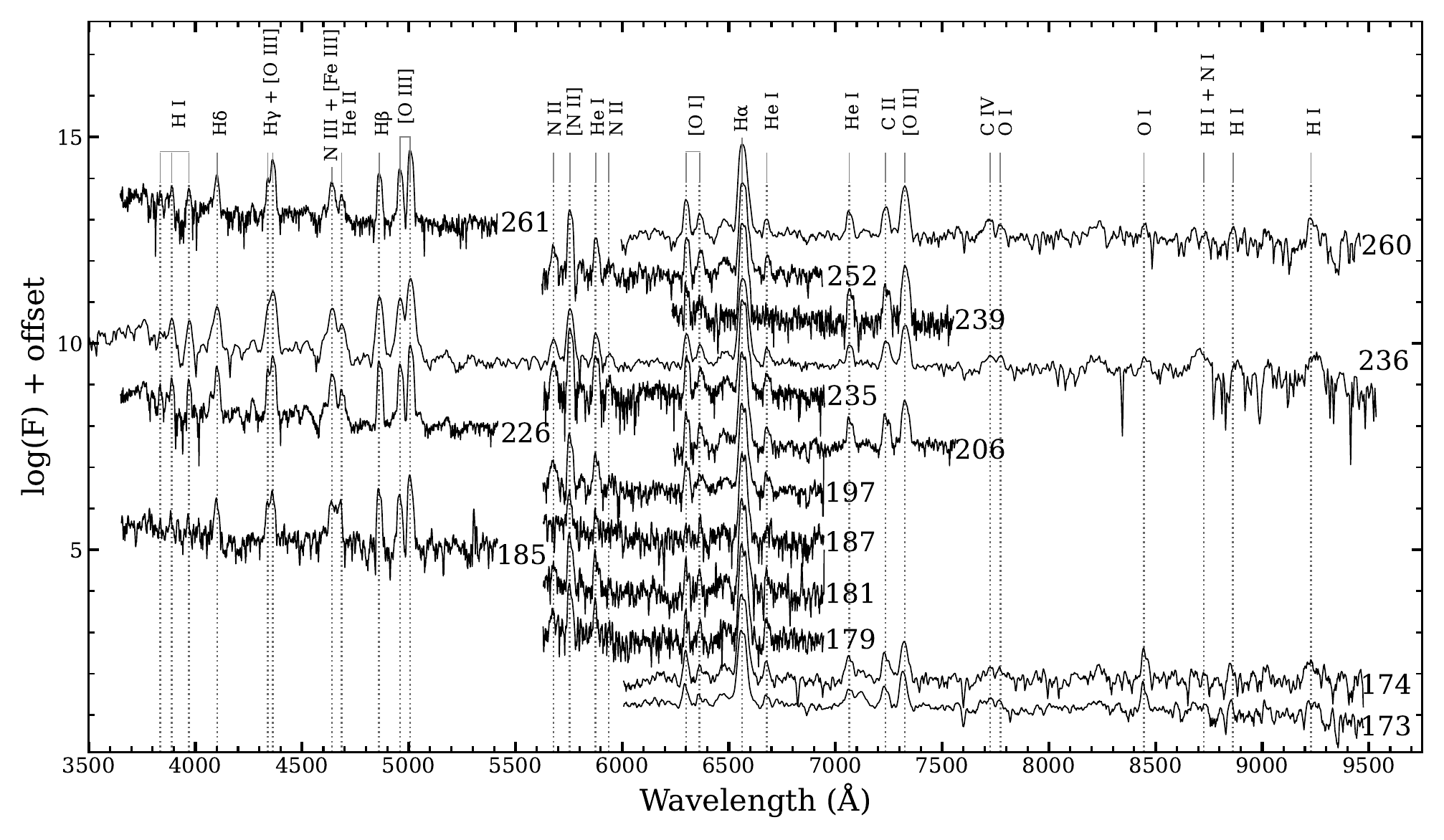}
\caption{ Spectroscopic evolution of Nova LMCN 2009-05a during the nebular phase, from day 173 (2009 October 25) to
day 261 (2010 January 21). The spectra prominently display strong forbidden lines, including [O III] at 4363, 4959, and 5007 \AA, and [N II] at 5755 \AA. }
	\label{spectral3}
\end{figure*}

\subsection{Physical parameters}
The physical parameters, e.g., optical depth, electron temperature, density, and mass, can be estimated from the dereddened line fluxes of oxygen and hydrogen.

The optical depth of oxygen and the electron temperature are estimated using the formulas found in \citet{wil94}, 
\begin{align}
\label{op_dep}
\dfrac{F_{\lambda6300}}{F_{\lambda6364}} = \dfrac{(1 - e^{-\tau})}{(1-e^{-\tau/3})}.
\end{align}
Using the value of $\tau$, the electron temperature (K) is then calculated from the intensity ratio of [O I] lines using the following relation, which is a derivation from equation (\ref{op_dep}) in \citet{wil94},

\begin{align}
\label{Te}
T_e =   \dfrac{11200}{ log [\dfrac{(43\tau)}{(1-e^{-\tau})} \times \dfrac{F_{\lambda6300}}{F_{\lambda5577}}]} 
\end{align} 
where F$_{\lambda5577}$, F$_{\lambda6300}$ and F$_{\lambda6364}$ are the line intensities of the [O I] 5577, 6300 and 6364 \AA\ lines. The electron temperature was estimated to be around 4520 K for day 79 which is typically seen in novae \citep{wil94}. 
However, for the subsequent epochs (days 171, 236, and 260), a similar temperature estimate was not possible due to the weak [O I] 5577 \AA{} line. For these later epochs, we estimated the electron density by assuming a typical nebular temperature of T$_e$ $\sim$ 10$^4$ K \citep{2005A&A...435.1031M}, using the [O III] 4363, 4959 and 5007 \AA{} lines in the following relation from \citet{ost06} 
\begin{equation}
\frac{j_{4959}+j_{5007}}{j_{4363}} = 7.9 \frac{e^{3.29\times10^{4}/T_{e}}}{1+4.5\times10^{-4}\frac{N_{e}}{T_{e}^{1/2}}}
\end{equation}
The electron density was not calculated for day 79 as the nebular [O III] lines were absent in the spectrum.

The oxygen mass can be estimated using the relation of \citet{wil94},
 \begin{equation}
\dfrac{m(O)}{M_{\odot}} = 152 d_{kpc}^{2} exp[\dfrac{22850}{T_e}] \times 10^{1.05E(B-V)}\frac{\tau}{1 - e^{-\tau}} F_{\lambda6300}
\end{equation}

The measured line flux ratios, optical depth, and the derived temperature, density, and mass are presented in Table \ref{phy_table}. 
The observed [O I] 6300/6364 line ratio (Table \ref{phy_table}, Col. 2) is approximately equal to the theoretical value of $\sim3$, which is expected for optically thin ejecta \citep{wil94}. The electron density $N_e$ was estimated to be of the order of $10^7$ cm$^{-3}$. 
The hydrogen mass can also be estimated using the relation by \cite{ost06},
\begin{align}
	\dfrac{m(H)}{M_{\odot}} = d^{2} \times 2.455 \times 10^{-2} \times \dfrac{I(H\beta)}{\alpha_{eff}N_e}
\end{align}
 where $\alpha_{eff}$ is the effective recombination coefficient obtained from \cite{hum95} and I(H$\beta$) is the flux of H$\beta$ line. The hydrogen mass is estimated to be 3.42 $\times$ 10$^{-6}$ M$_{\odot}$, calculated as the mean value of two epochs (Table \ref{phy_table}, Col. 7).
Here we have used the distance adopted in Section \ref{reddening and distance} from \cite{pietrzynski2013eclipsing}.

\begin{table*}
\centering
\caption{The [O I] line flux ratio, the corresponding opacity $\tau_{6300}$, $T_e$, and mass derived for nova at each epoch according to \cite{wil94}. The epochs (days 79, 171, and 236) were selected as full optical spectra were available for these epochs. See text for more details.}
\begin{tabular}{@{}lcccccc@{}}
\toprule
Obs. date (days) & $r = \frac{F_{6300}}{F_{6364}}$ & $\tau_{6300}$ & $T_e$ (K) & $N_e$ & $M_{\text{OI}} (M_\odot)$   & $M_{\text{H}} (M_\odot)$  \\ \midrule
23/07/09 (+79) & 2.98 & 0.01           & 4520    & ...                    & $5.85 \times 10^{-7}$  & ...\\
23/10/09 (+171) & 2.37 & 0.78          & 1e4     &  $1.40 \times 10^{7}$  & $2.00 \times 10^{-7}$  & $4.17 \times 10^{-6}$\\
27/12/09 (+236) & 2.68 & 0.35          & 1e4     &  $1.60 \times 10^{7}$  & $1.93 \times 10^{-7}$  & $2.66 \times 10^{-6}$\\
20/01/10 (+260) & 2.65 & 0.37          & 1e4     &  $2.08 \times 10^{7}$  & $1.78 \times 10^{-7}$  & ... \\ \bottomrule
\label{phy_table}
\end{tabular}
\end{table*}

\begin{figure}
    \centering
    \includegraphics[width=1.0\linewidth]{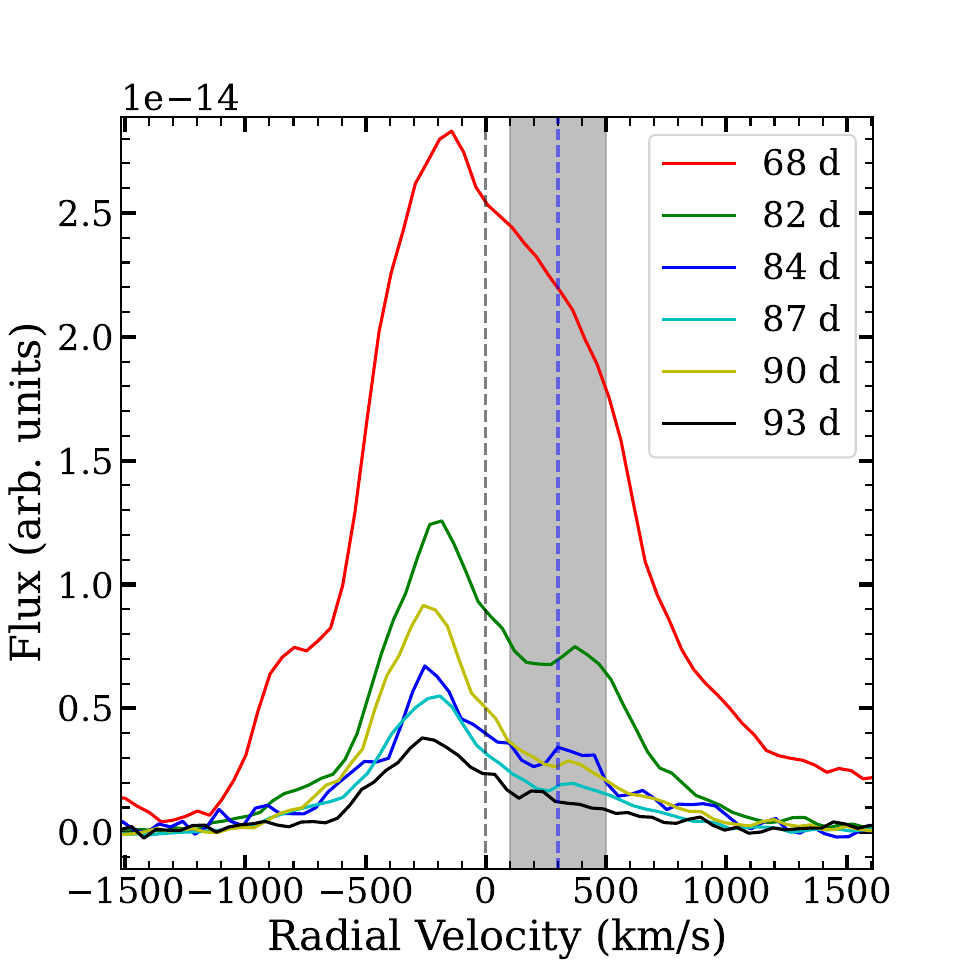}
    \caption{Evolution of the H$\alpha$ line profiles in the nova spectrum from 68 to 93 days after outburst. All profiles are plotted in velocity space, corrected for the LMC velocity (278 $\kms$ \cite{1987A&A...171...33R}). The vertical dashed black line marks the zero velocity. A discrete absorption feature is highlighted by a shaded region, centered at approximately +300 $\kms$ and marked by a vertical dashed blue line, corresponding to the disappearance of the redward part of the profile near the onset of the photometric minimum.
}
    \label{radial}
\end{figure}

\begin{figure*}
\centering
	\includegraphics[scale=0.39]{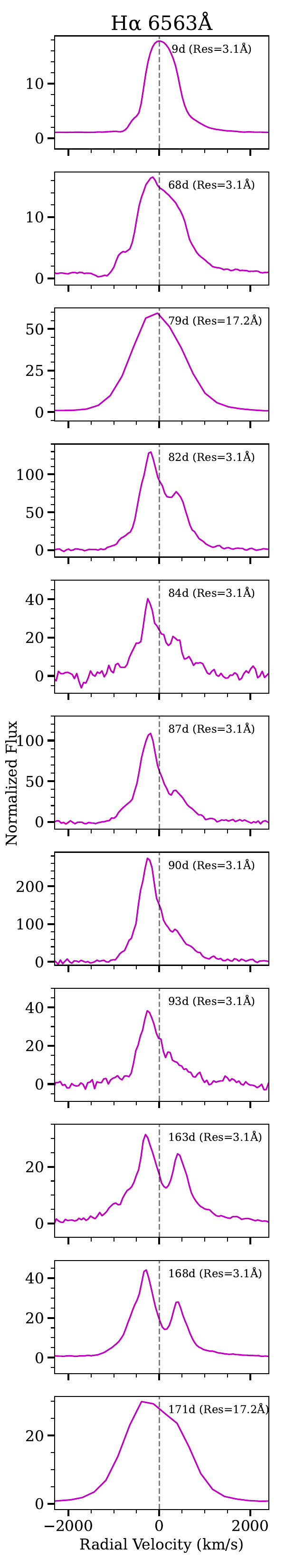}
    \includegraphics[scale=0.39]{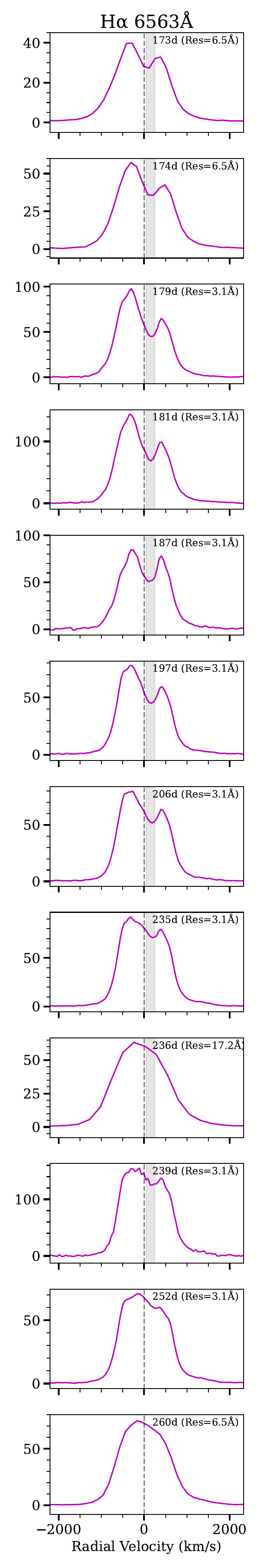}
    \includegraphics[scale=0.39]{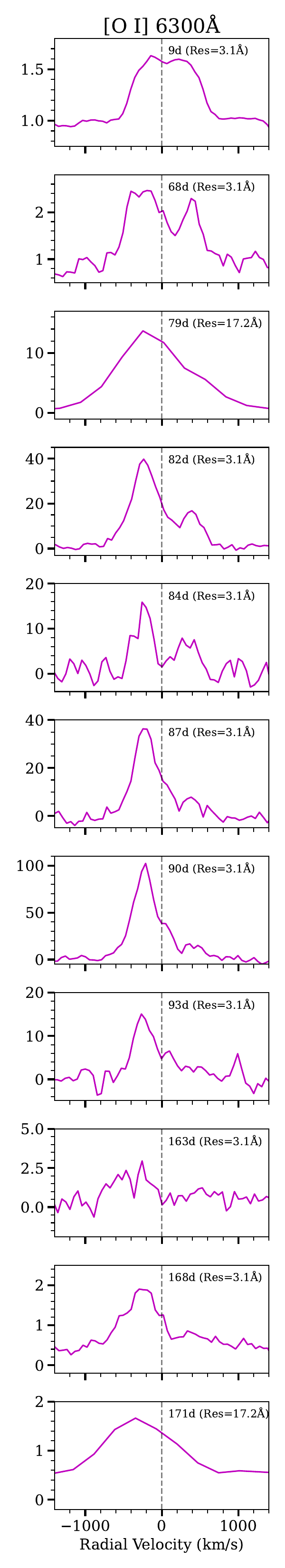}
    \includegraphics[scale=0.39]{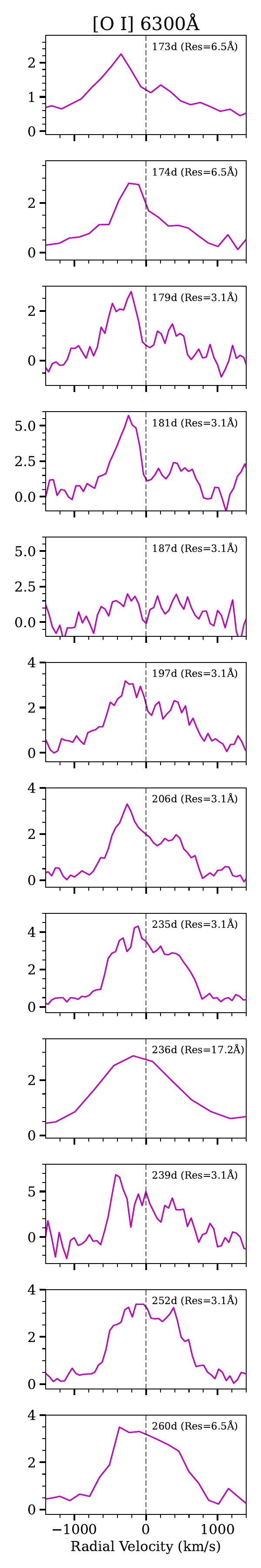}
\caption{H $\alpha$ and [O I] 6300 \AA{} line profiles with radial velocities corrected for the LMC velocity (278 $\kms$ \cite{1987A&A...171...33R}). The shaded regions mark the location of the reversal feature observed during the nebular phase  (more details in section \ref{Spectral evolution}).}
	\label{h alpha}
\end{figure*}

\subsection{Photoionization Modeling}\label{cloudy}
We employed the photoionization code CLOUDY (version C23.01; \cite{chatzikos20232023}) to model the early-decline (day 79) and the nebular-phase (day 236) optical spectrum of LMCN 2009-05a. This modeling aimed to investigate the physical conditions and elemental abundances of the ejecta. By selecting the "pre-dust" phase (79 days) and the "nebular phase" (236 days), we focused on a period with minimal dust interference, enabling a more accurate examination of elemental emissions and their contributions to the observed spectra. CLOUDY, a photoionization code, utilizes detailed microphysics to simulate the physical conditions of non-equilibrium gas clouds under the influence of an external radiation field. It generates emission-line spectra based on specified parameters such as the density, temperature, and chemical composition of the gas cloud.

In this study, we considered a central ionizing source with a blackbody spectrum characterized by a temperature \( T_{\text{BB}} \) (K) and luminosity \( L \) (erg s\(^{-1}\)), surrounded by a spherically symmetric ejecta. The ejecta characteristics include its density, inner and outer radii, covering factor, filling factor, and elemental composition. The density of the ejecta is determined by the total hydrogen number density, \( n(r) \) (cm\(^{-3}\)).
In our model, the hydrogen density \( n(r) \) and the filling factor \( f(r) \) vary with radius following the relations:  

\begin{equation}  
    n(r) = n(r_{\text{in}}) \left( \frac{r}{r_{\text{in}}} \right)^{\alpha}  
\end{equation}  

\begin{equation}  
    f(r) = f(r_{\text{in}}) \left( \frac{r}{r_{\text{in}}} \right)^{\beta}  
\end{equation}  

where \( r_{\text{in}} \) represents the inner radius of the ejecta, \( \alpha \) and \( \beta \) are the power-law exponents governing the radial dependence of density and filling factor, respectively.
In this study, we adopted $\alpha = -3$, representing ballistic expansion of the ejecta, a typical filling factor of 0.1, and $\beta = 0$, in alignment with values employed in previous nova studies (e.g., \cite{2010AJHelton,raj2018cloudy2676,bisht2025spectrophotometric}, and references therein). The chemical composition of the ejecta was defined using the abundance parameter in \textsc{Cloudy}.
We aimed to estimate the elemental abundances of those elements for which emission lines were found in the observed spectra while keeping the abundances of other elements fixed at their solar values \citep{2010Ap&SSGrevesse}.
Previous photoionization modeling studies (e.g., \cite{shore2003early,2010AJHelton,raj2018cloudy2676,2020MNRASPavana,raj2024dustyaftermathrapidnova}, and references therein) have demonstrated that a two-component modeling approach yields a more physically consistent representation of nova ejecta. Novae typically exhibit an inhomogeneous density structure, characterized by clumpy and diffuse regions within the expanding material \citep{paresce1995structure,BodeEvansBook2008,williams2013novae,chomiuk2021new}. In this study, we considered a two-component density structure to closely match the observed fluxes.
The observed emission lines were dereddened using the E(B-V) value determined in Section \ref{reddening and distance}.
A more detailed description of the modeling process followed here can be found in \citet{raj2018cloudy2676,raj2024dustyaftermathrapidnova,bisht2025spectrophotometric}.

We evaluated the goodness of fit to determine the best-fit model using the $\chi^2$ and reduced $\chi^2$ tests. A total of 27 emission lines from the day 79 spectrum and 22 lines from the day 236 spectrum were used for the fitting. Tables \ref{d79_flux_table} and \ref{d236_flux_table} present the dereddened relative fluxes of the observed and best-fit model lines, along with their corresponding $\chi^2$ values. The derived physical parameters and elemental abundances are provided in Table \ref{cloudy_result}. The best-fit synthetic spectra, generated using the derived model parameters, are shown alongside the observed spectra in Figures~\ref{cloudy79}
and~\ref{cloudy236} for both epochs.

On day 79, the nova was in the Fe II phase \citep{2024MNRAS.527.9303A}, with all the Fe II and [O I] lines originating from a high-density region with a density of \(2.5 \times 10^{11} \,\text{cm}^{-3}\). By day 236, the nova had transitioned to the nebular phase, with helium and nitrogen lines indicating that it had undergone a later He/N phase (see Section \ref{Spectral evolution}). The [O I] lines were produced in a region with a density of \(2.0 \times 10^{8} \,\text{cm}^{-3}\), while the [O III] and N II lines originated from a relatively lower-density region. The He I and He II lines originate from both the high- and low-density regions. 
Our modeling reproduced the Fe II lines on day 79 and the He and N lines on day 236, both having nearly constant luminosity. The observed spectral evolution is attributed to changes in ionization temperature and density within the ejecta, consistent with the findings of \citep{2012BASI...40..185S,2024MNRAS.527.9303A}. The estimated abundance values show that nitrogen and oxygen are higher than the solar values, while calcium is about half the solar value.

\begin{figure*}
    \centering
    \includegraphics[width=1.0\linewidth]{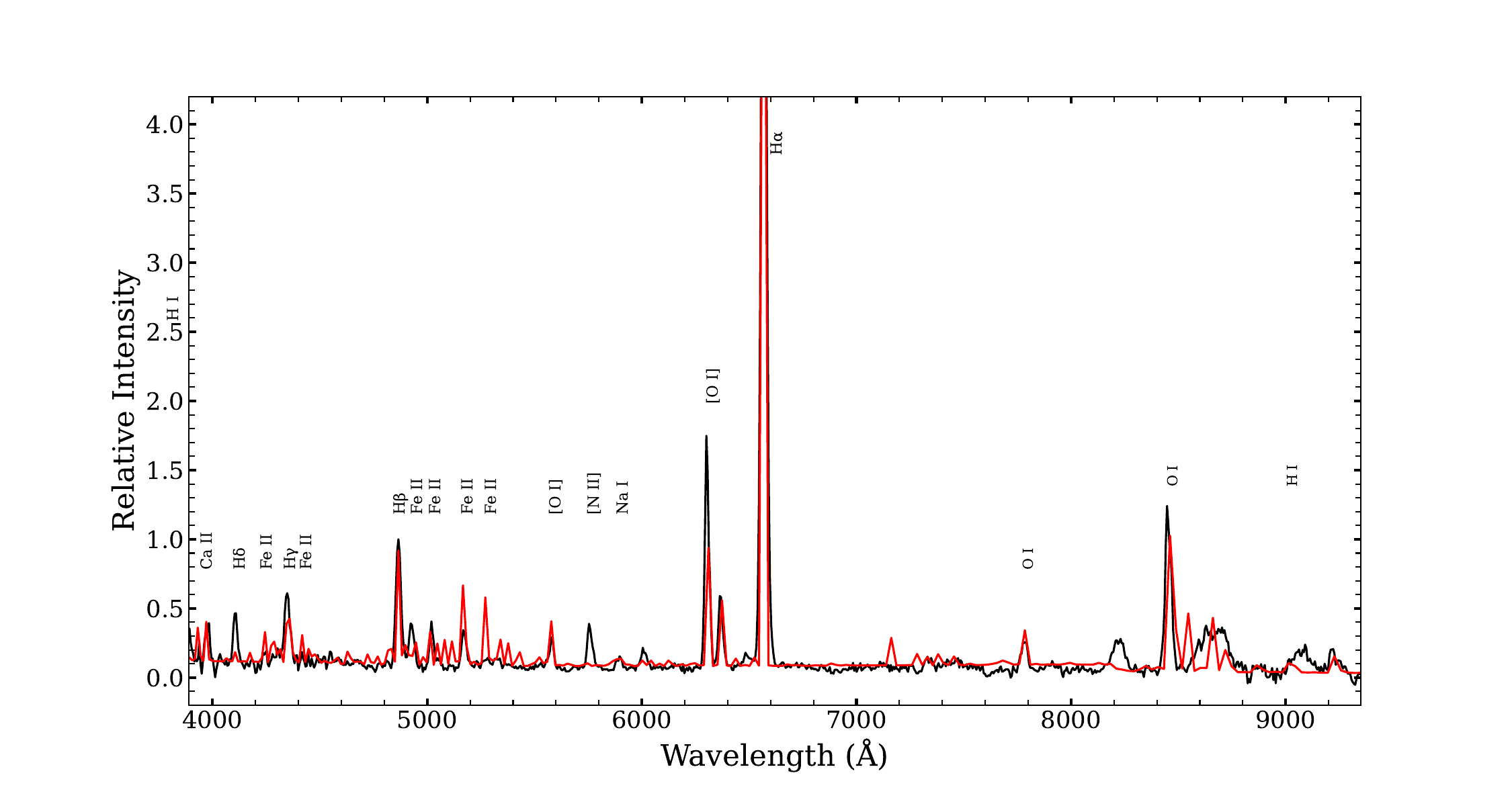}
    \caption{ Observed optical spectrum (black) of Nova LMCN 2009-05a on 23 July 2009 (day 79), plotted over with the best-fitting CLOUDY model (red). }
    \label{cloudy79}
\end{figure*}

\begin{figure*}
    \centering
    \includegraphics[width=1.0\linewidth]{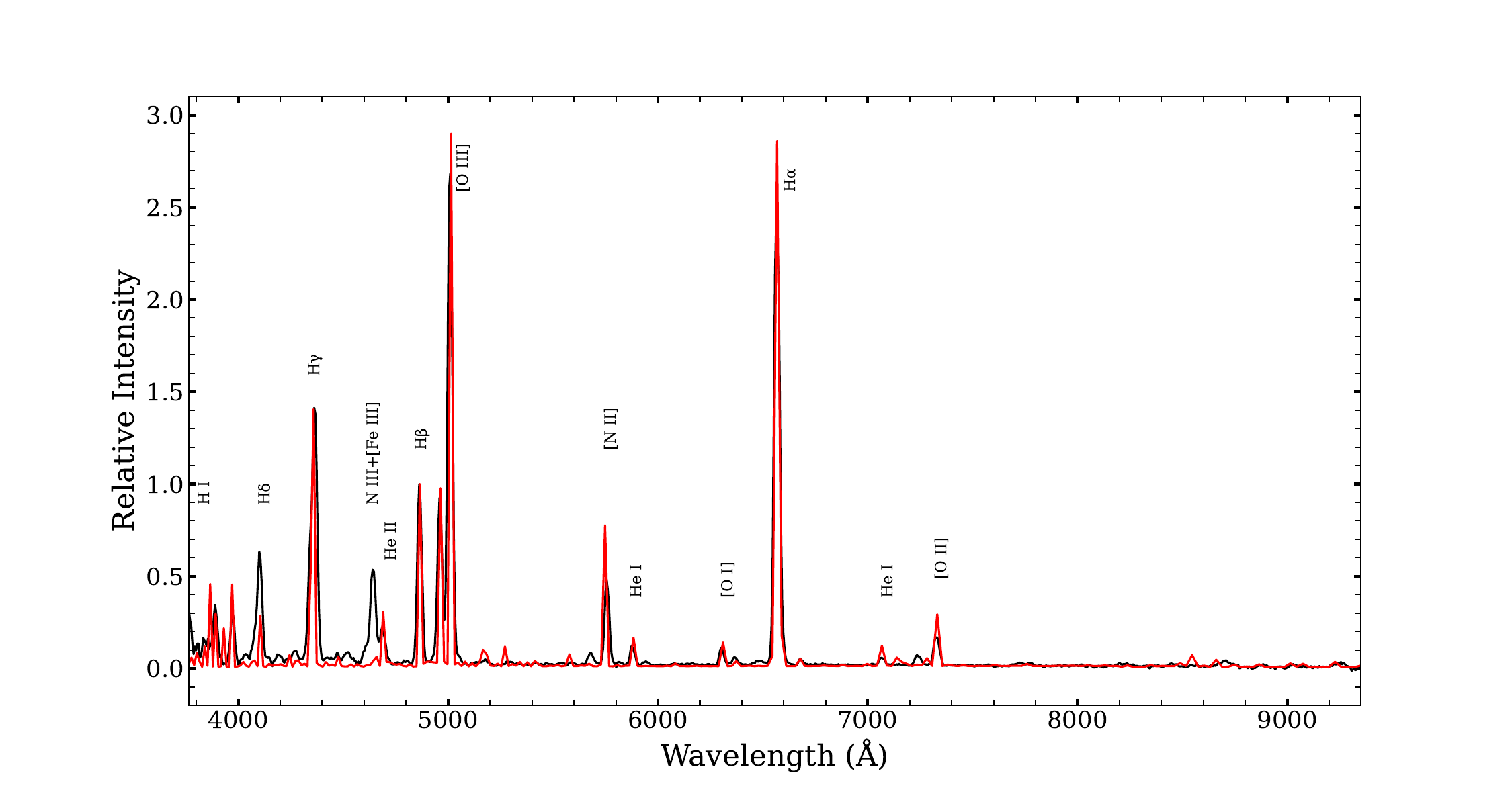}
    \caption{Observed optical spectrum (black) of Nova LMCN 2009-05a on 2009 December 27 (day 236), plotted over with the best-fitting CLOUDY model (red).}
    \label{cloudy236}
\end{figure*}

\begin{table}
	\caption{Observed and best-fit optical \texttt{CLOUDY} model line flux ratios on day 79 for LMCN 2009-05a}
	\label{d79_flux_table}
	\begin{center}
		\resizebox{\hsize}{!}{%
			\begin{tabular}{lclcc}
				\hline
				\hline
				\textbf{Line ID} & \boldmath{$\lambda$ (\AA)} & \textbf{Observed}$^{a}$ & \textbf{Modelled}$^{a}$ & \boldmath{$\chi^2$}  \\
				\hline
			    Ca II           & 3934       & 8.98E$-$02 & 2.33E$-$01 & 3.31E$-$01 \\
				Ca II+H I       & 3968       & 4.67E$-$01 & 3.25E$-$01 & 2.26E$-$01 \\
				H I             & 4102       & 3.41E$-$01 & 1.51E$-$01 & 5.84E$-$01 \\
				Fe II           & 4179       & 7.75E$-$02 & 6.78E$-$02 & 1.04E$-$03 \\
				Fe II           & 4233       & 1.19E$-$01 & 1.72E$-$01 & 4.39E$-$02 \\
				Fe II           & 4276       & 8.21E$-$02 & 1.22E$-$01 & 1.79E$-$02 \\
				Fe II           & 4303       & 5.60E$-$02 & 5.29E$-$02 & 1.48E$-$04 \\
                H I             & 4340       & 5.87E$-$01 & 3.12E$-$01 & 1.21E$+$00 \\
				  Fe II           & 4385       & 4.60E$-$02 & 1.50E$-$02 & 1.53E$-$01 \\
				Fe II           & 4417       & 5.44E$-$02 & 5.76E$-$02 & 2.67E$-$04 \\
                Fe II           & 4584       & 2.84E$-$02 & 3.55E$-$02 & 8.25E$-$04 \\
				  H I             & 4861       & 1.00E$+$00 & 1.00E$+$00 & 0.00E$+$00 \\
				  Fe II           & 4924       & 2.87E$-$01 & 2.27E$-$01 & 3.91E$-$02 \\
                Fe II           & 5018       & 2.32E$-$01 & 1.45E$-$01 & 1.20E$-$01 \\  
                Fe II           & 5169       & 2.84E$-$01 & 2.03E$-$01 & 1.61E$-$01 \\
                Fe II           & 5265       & 1.06E$-$01 & 1.62E$-$01 & 5.02E$-$02 \\
                Fe II           & 5334       & 1.02E$-$01 & 1.76E$-$01 & 8.61E$-$02 \\
                Fe II           & 5363       & 4.43E$-$02 & 1.24E$-$02 & 1.63E$-$02 \\
				  {[}O I{]}       & 5577       & 2.07E$-$01 & 2.31E$-$01 & 1.45E$-$02 \\
                {[}N II{]}      & 5755       & 2.94E$-$01 & 1.83E$-$01 & 3.02E$-$01 \\
                Na I            & 5892       & 1.20E$-$01 & 1.02E$-$01 & 8.20E$-$03 \\
				{[}O I{]}       & 6300       & 1.39E$+$01 & 1.15E$+$01 & 9.47E$-$01 \\
				{[}O I{]}       & 6364       & 4.73E$-$01 & 3.94E$-$01 & 9.88E$-$02 \\
				H I             & 6563       & 8.82E$+$00 & 9.80E$+$00 & 15.64E$+$00 \\
				O I             & 7773       & 3.04E$-$01 & 1.25E$-$01 & 5.14E$-$01 \\
                O I             & 8446       & 1.54E$+$01 & 1.87E$+$01 &1.24E$+$00 \\
				Ca II + H I     & 8665       & 1.62E$+$01 & 1.00E$+$01 & 4.31E$+$00 \\
				\hline
		    \end{tabular}}
	\end{center}
	$^{a}$Relative to H$\beta$
\end{table}

\begin{table}
	\caption{ Observed and best-fit optical CLOUDY model line
flux ratios on day 236 for LMCN 2009-05a}
	\label{d236_flux_table}
	\begin{center}
		\resizebox{\hsize}{!}{%
			\begin{tabular}{lclcc}
				\hline
				\hline
				\textbf{Line ID} & \boldmath{$\lambda$ (\AA)} & \textbf{Observed}$^{a}$ & \textbf{Modelled}$^{a}$ & \boldmath{$\chi^2$}  \\
				\hline
			    H I           & 3835       & 1.15E$-$01 & 8.54E$-$02 & 2.31E$-$02 \\
				  H I           & 3889       & 2.82E$-$01 & 1.21E$-$01 & 6.49E$-$01 \\
				  H I           & 3970       & 2.25E$-$01 & 1.76E$-$01 & 5.91E$-$02 \\
				H I  & 4102    & 6.63E$-$01 & 3.55E$-$01 & 2.37E$+$00 \\
                H I           & 4340       & 5.40E$-$01 & 4.49E$-$01 & 6.18E$-$02 \\
                {[}O III{]}   & 4363       & 1.40E$+$00 & 1.45E$+$00 & 6.17E$-$02 \\
				  N III + {[}Fe III{]} & 4640& 6.54E$-$01 & 1.15E$-$01 & 7.26E$+$00 \\
				He II         & 4686       & 2.49E$-$01 & 3.22E$-$01 & 1.32E$-$01 \\
				H I           & 4861       & 1.00E$+$00 & 1.00E$+$00 & 0.00E$+$00 \\
				  {[}O III{]}   & 4959       & 1.03E$+$00 & 9.78E$-$01 & 6.71E$-$02 \\
				  {[}O III{]}   & 5007       & 2.94E$+$00 & 2.91E$+$00 & 1.88E$-$02 \\
				Mg I          & 5172       & 6.99E$-$02 & 6.24E$-$03 & 1.01E$-$01 \\
				  N II          & 5679       & 7.14E$-$02 & 1.60E$-$02 & 7.69E$-$02 \\
				{[}N II{]}    & 5755       & 4.80E$-$01 & 6.91E$-$01 & 1.22E$+$00 \\
				He I          & 5876       & 1.15E$-$01 & 2.22E$-$01 & 2.82E$-$01 \\
				{[}O I{]}     & 6300       & 9.63$-$02 & 9.53E$-$02 & 4.40E$-$05 \\
				{[}O I{]}     & 6364       & 4.71E$-$02 & 3.04E$-$02 & 6.97E$-$03 \\
				H I           & 6563       & 3.12E$+$00 & 2.93E$+$00 & 8.43E$-$01 \\
				He I          & 6678       & 2.82E$-$02 & 3.75E$-$02 & 2.19E$-$03 \\
                He I          & 7065       & 5.04E$-$02 & 1.08E$-$01 & 8.29E$-$02 \\
				{[}Ar IV{]}  & 7237       & 8.34E$-$02 & 3.71E$-$03 & 1.59E$-$01 \\
                {[}O II{]}    & 7325       & 2.11E$-$01 & 2.99E$-$01 & 1.92E$-$01 \\
				\hline
		    \end{tabular}}
	\end{center}
	$^{a}$Relative to H$\beta$
\end{table}

\begin{table}
	\caption{Best-fit Optical CLOUDY model parameters obtained on day 79 and 236 for nova LMCN 2009-05a}
	\label{cloudy_result}
	\begin{center}
		\resizebox{\hsize}{!}{%
			\begin{tabular}{l c l } 
				\hline\hline
				\textbf{Parameter} & \textbf{Day 79}  & \textbf{Day 236} \\ [0.5ex] 
				\hline 
				T$_{BB}$ ($\times$ 10$^{4}$ K) & 1.61 $\pm$ 0.15& 19.9 $\pm$ 0.10 \\ [0.25ex]
				Luminosity ($\times$ 10$^{36}$ erg/s) & 6.83 $\pm$ 0.3  &7.94 $\pm$ 0.08\\ [0.25ex]
				Clump Hydrogen density ($ cm^{-3}$) & 2.51$\times$ 10$^{11}$ & 1.99$\times$ 10$^{8}$ \\ [0.25ex]
				Diffuse Hydrogen density ($ cm^{-3}$) & 2.24$\times$ 10$^{8}$ &1.58$\times$ 10$^{7}$\\ [0.25ex]
				Covering factor (clump) & 0.70 & 0.45\\ [0.25ex]
				Covering factor (diffuse) & 0.30 & 0.55\\ [0.25ex]
				$\alpha$ & -3.00 & -3.00\\ [0.25ex]
				Inner radius ($\times$ 10$^{14}$ cm) & 7.22 &22.6 \\ [0.25ex]
				Outer radius ($\times$ 10$^{15}$ cm) & 1.16 & 3.46\\ [0.25ex]
				Filling factor  & 0.1 & 0.1\\ [0.25ex]
                He/He$_{\odot}$ & 1.0  & 1.35 $\pm$ 0.2 (4)\\ [0.25ex]
                N/N$_{\odot}$ & 75.00 $\pm$ 15 (1)$^{a}$ &  45.00 $\pm$ 6 (2)\\ [0.25ex]
				O/O$_{\odot}$ & 10.0 $\pm$ 2.0 (5) &  4.16 $\pm$ 0.8 (7)\\ [0.25ex]
                Na/Na$_{\odot}$ & 1.5 $\pm$ 0.2 (1) &  1.0\\ [0.25ex]
                Ca/Ca$_{\odot}$ & 0.5 $\pm$ 0.2 (2) & 0.5 \\ [0.25ex]
                Fe/Fe$_{\odot}$ & 2.0 $\pm$ 0.4 (13) & 1.0\\ [0.25ex]
				Number of observed lines (n) & 27 & 22\\ [0.25ex]
				Number of free parameters (n$_{p}$) & 10 & 11\\ [0.25ex]
				Degrees of freedom ($\nu$) & 17 & 11\\ [0.25ex]
				Total $\chi^{2}$ & 26.00 & 13.68\\ [0.25ex]
				$\chi^{2}_{red}$ & 1.53 & 1.24\\ [0.25ex]
				\hline
		\end{tabular}}
	\end{center}
	$^{a}$The number of lines available to obtain abundance estimate is as shown in the parenthesis.
\end{table}

\subsection{Dust mass and temperature}\label{dust masss and temp}
A sudden decrease in brightness in the optical light curve indicated that the onset of a dust dip between 67 and 89 days after the outburst. Additionally, an increase in the fluxes of the NIR band during dust formation (Fig. \ref{lc_nir}) clearly indicate substantial dust formation. The observations from the 
WISE \citep{wright2010wide} in 3.4 (W1), 4.6 (W2), 12 (W3) and 22 ~$\mu$m (W4) bands also support emission from the dust at longer wavelengths (see Fig. \ref{wise_image}) around day 395. We estimated the temperature of the dust shell at around 395 days since discovery using the magnitudes from WISE as 700 ± 50 K. However, the temperature estimate for the dust shell may have a large uncertainty, as we assume that the isothermal dust and we have used only four wavelengths to
fit the SED. The SED has peak emission near W1 (3.4 µm) in fig. \ref{sed}. We estimate the dust mass following \cite{evans2017rise}, assuming that the grains are spherical and that the dust is composed of carbonaceous material. Using the relations given by \cite{evans2017rise}, we find that the dust masses for the grains of amorphous carbon (AC) and graphitic carbon (GR) are as follows. 

For optically thin amorphous carbon,
\begin{equation}
\frac{M_{dust,AC}}{M_{\odot}} \simeq 4.65 \times 10^{19} \frac{(\lambda f_{\lambda})_{max}}{T^{4.754}_{dust}}
\end{equation}
and for optically thin graphitic carbon,
\begin{equation}
\frac{M_{dust,GR}}{M_{\odot}} \simeq 4.14 \times 10^{21} \frac{(\lambda f_{\lambda})_{max}}{T^{5.315}_{dust}}
\end{equation} 
where we assume the distance $D$ = 50 kpc, the density of the carbon grains $\rho$ = 2.25 gm cm$^{-3}$, and ($\lambda$ $f_{\lambda}$)max is in unit of W m$^{-2}$. The dust mass, which is independent of grain size
\citep{evans2017rise}, is estimated to be $\sim$ 2.24$\times 10^{-9}M_{\odot}$  and
5.05$\times 10^{-9}M_{\odot}$ , for AC and GR grains, respectively.

\subsection{Grain Size}\label{size}
For grain size using the relation given by \cite{gehrz2018temporal},

\begin{equation}
a \simeq \frac{L_0}{16 \pi R^2 A \sigma T^{\beta + 4}}
\end{equation}
where \( L_0 \) is the bolometric luminosity of the nova. On day 395, we have 
\( T = 700 \, \text{K} \), and with \( L_0 = 4.64 \times 10^4 L_{\odot} \). The expansion velocity of the ejecta \( V_0 = 690~\text{km s}^{-1} \), estimated from the HWHM of emission lines in our spectra, following the method used by \citet{evans2017rise}. The dust shell radius \( R = 2.59 \times 10^{15}\ \, \text{cm} \) on day 395 is calculated using \( R = V_0 \times t \), following the method described in \citet{gehrz2018temporal}.
We determined grain radius of \( a = 0.06 \pm 0.01\, \mu\text{m} \) for amorphous carbon and \( a = 0.13 \pm 0.04\, \mu\text{m} \) for graphite carbon, assuming the nova maintained a constant bolometric luminosity during this period. These values are consistent with typical grain sizes found in novae at later stages of dust evolution \citet{gehrz2018temporal}.

\begin{figure}
\centering
	\includegraphics[scale=0.35]{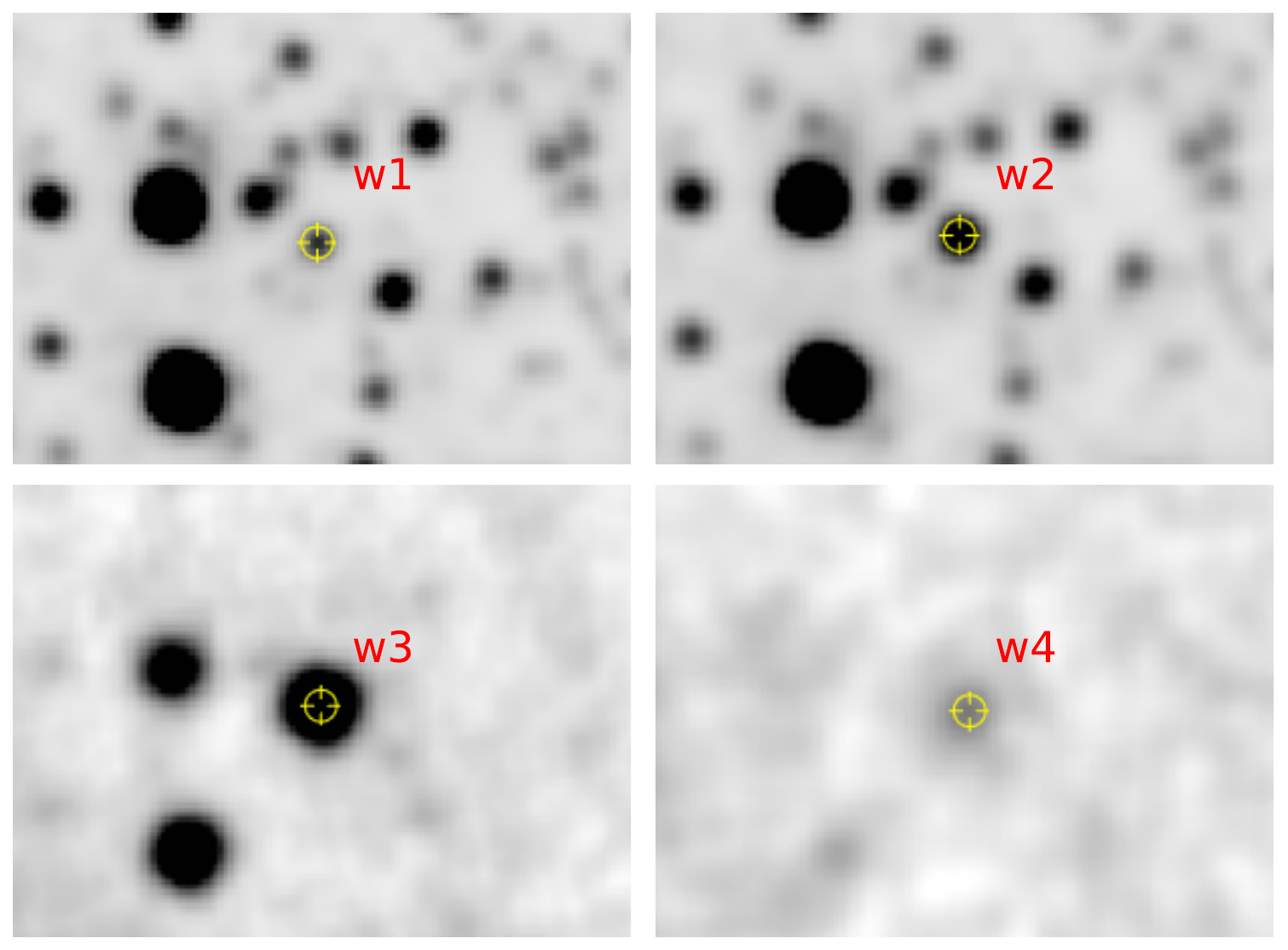}
	\caption{A mosaic of a 3x3 arc minute square field around LMCN 2009-05a. The source is detected in all 4 WISE bands: W1 (3.4~$\mu$m), W2 (4.6~$\mu$m), W3 (12~$\mu$m) and W4 (22~$\mu$m); the emission at W2 and W3 bands is pronounced. 
    The WISE images were obtained from the WISE portal and represent coadded data from observations taken between January 2010 and January 2011(more details in section \ref{dust masss and temp})}.
	\label{wise_image}
\end{figure}

\begin{figure}
    \centering
    \includegraphics[width=1.0\linewidth]{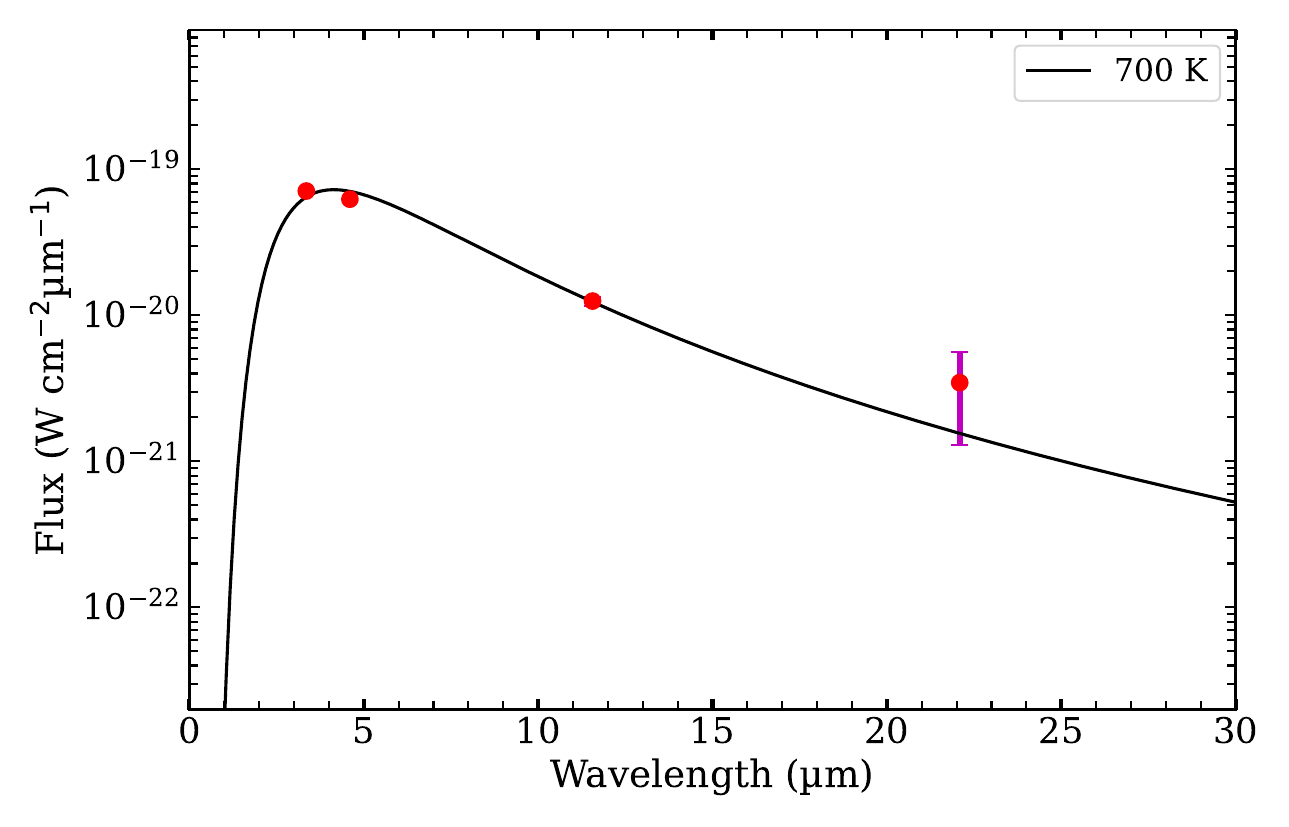}
    \caption{The SED shows a best fit blackbody to the WISE data taken on day 395, with a temperature of about 700
K.}
    \label{sed}
\end{figure}

\section{Discussion}\label{Discussion}
The spectrophotometric evolution of LMCN 2009-05a shows that it is a classical moderately fast nova. This nova is marked by notable dust formation and changes in ionization structure. It fits the D-class light curve morphology according to the classification by \cite{strope2010catalog}. This is confirmed by the clear optical dust dip and the complementary NIR enhancements.

The estimated absolute magnitude  and the outburst luminosity  place LMCN 2009-05a within the relatively faint group of classical novae. Together with the inferred white dwarf mass $\sim$ 0.77 $\pm$ 0.10M$_\odot$, this supports the idea of a CO white dwarf progenitor \citep{livio1992classical}. The presence of Fe II multiplets and a strong P-Cygni absorption in early stages confirms dense, slowly expanding ejecta typical of Fe II novae \citep{williams2012origin}.

Dust formation began 78 days after the outburst. This aligns with the well-established correlation between \(t_{\mathrm{cond}}\) and the decline time (\(t_2\)), first noted by \citet{shafter2011spitzer} and further discussed by \citet{williams2013rapid,2025arXiv250104098C}. We investigated this relationship for novae in the LMC, including LMC 1998\#1, LMCN 1999-09a, LMCN 2005-11a, LMCN 2009-05a, LMCN 2011-08a, N LMC 2013, and LMCN 2017-11a --- the only LMC novae currently known to have formed dust. Our findings confirm that the \(t_{\mathrm{cond}} - t_2\) relation holds for these LMC novae as well (see Fig.~\ref{t2td}), consistent with previous results in the Milky Way.
Further, we fitted a power-law to both the LMC novae (see Table~\ref{tcond_lmc}) and the Galactic novae (see Table~\ref{tcond}). For the combined sample, the best-fit relation is
\begin{equation}
t_{\mathrm{cond}} = 5.34 \times t_2^{0.67}
\label{complete_fit}
\end{equation}
within the \(1\sigma\) confidence level (see Fig.~\ref{powerfit}). 
For the LMC novae alone, the best-fit relation is
\begin{equation}
t_{\mathrm{cond}} = 5.49 \times t_2^{0.65}
\label{lmc_fit}
\end{equation}
For the relation derived for the LMC novae alone, the slopes and coefficients are very similar to those of the combined fit, indicating that the dust formation timescale–decline rate correlation is consistent across both Galactic and LMC environments. This suggests that the dominant physical processes governing dust formation are similar in both environments, despite differences in metallicity and stellar population properties. Our empirical relations given in Equations \ref{complete_fit} and \ref{lmc_fit} are independent of the dust grain type. 
The dashed green and blue lines in Figure~\ref{powerfit} represent the graphite and ACH2 grain model fits derived by \cite{williams2013rapid}. The close agreement between these model curves and the combined nova sample shows that the \cite{williams2013rapid} models are broadly valid across different galactic environments. Overall, the continued presence of the \(t_{\mathrm{cond}} - t_2\) relationship in different galactic environments highlights its reliability and the fundamental physical processes governing dust formation.

The asymmetric line profiles in the H$\alpha$ and [O I] lines during the dust dip, which occurs near 82 day, indicate selective obscuration of the receding ejecta by dust. This fits models that suggest dust forms preferentially in the nova's equatorial plane \citep{gehrz2018temporal,shore2018spectroscopic}. The following profile changes in the nebular phase may indicate geometric changes or expansion that exposes deeper layers of the shell.

Photoionization modeling using CLOUDY shows a shift from a dense, clumpy region on day 79 to a more diffuse nebular-phase ejecta on day 236. There are notable increases in nitrogen and oxygen levels compared to solar, suggesting that material processed by the CNO cycle was ejected. The continued nearly constant ionizing luminosity between these times implies a slow change in optical depth instead of a rapid drop in the central source.

The dust shell parameters indicate dust temperatures around 700 K, with grain sizes \(0.06 \pm 0.01\,\mu\text{m}\) (AC) and \(0.13 \pm 0.04\,\mu\text{m}\) (GR), which match optically thin carbonaceous dust. The dust masses ($\sim$ $10^{-9}\,M_{\odot}$) fall within the usual range observed in other dusty novae \citep{evans2017rise,raj2024dustyaftermathrapidnova}. 
The dust mass and grain sizes estimated on day 395 post-outburst ($\sim$ 317 days after the onset of dust formation) suggest that the grains may have been destroyed afterward, possibly due to UV radiation or collisional sputtering as the nova progressed.

An interesting aspect of our observations is the inclination-dependent visibility of dust dips. Comparison with D-class novae \citep{strope2010catalog} shows a tendency for higher inclinations in systems where dust dips occur \citep{2021PhDT........46B}. This could be due to dust concentrated around the equator and the geometry of the line of sight. Although inclination data for LMC novae are limited, this hypothesis seems consistent with shock-dust scenarios.

\begin{table}
\centering
\caption{Dust formation timescales of LMC novae}
\label{tcond_lmc}
\resizebox{\hsize}{!}{%
\begin{tabular}{lccc}
\toprule
Nova & $t_2$ (days) & $t_{\text{cond}}$ (days) & References \\
\midrule
LMC 1988\#1    & 23 &  59     & 1 \\
LMCN 1999-09a  & 16 & 19 & 2 \\
LMCN 2005-11a  & 70.4 & 86.5 & 3 \\
LMCN 2009-05a  & 46 & 78  & 4 \\
LMCN 2011-08a  & 9 & 34 & 2 \\
N LMC 2013     & 52 & 48  & 3 \\
LMCN 2017-11a  & 121 & 125 & 5 \\

\bottomrule
\end{tabular}}
\vspace{0.5em}
\noindent \textbf{References.} (1) \cite{shafter2011spitzer}; (2) \cite{mroz2016ogle}; (3) \cite{2025arXiv250104098C}; (4) This paper; (5) \cite{2019arXiv190309232A}
\end{table}

\begin{figure}
    \centering
    \includegraphics[width=1.0\linewidth]{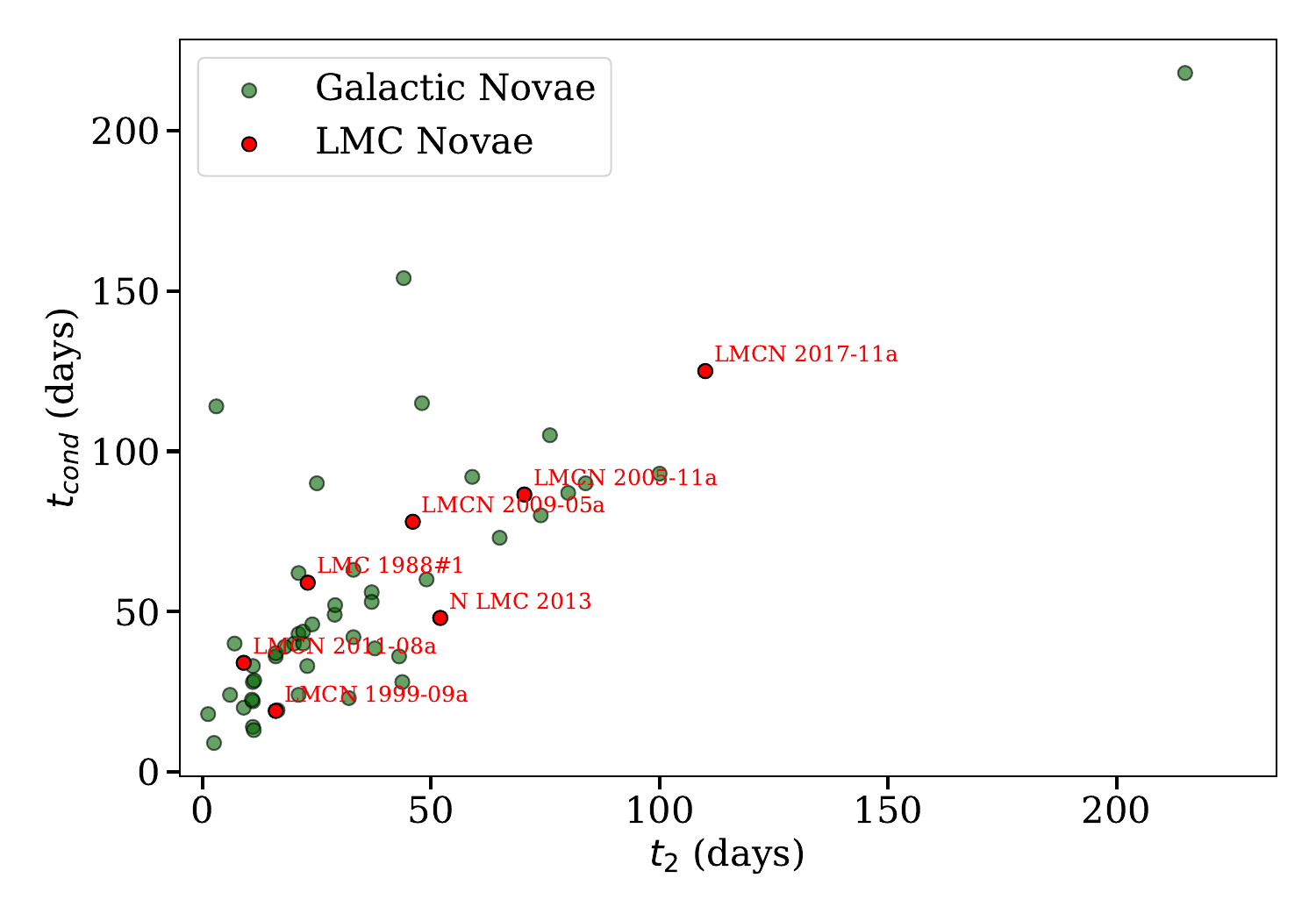}
    \caption{Correlation between the dust \( t_{\text{cond}} \) and \( t_2 \) for novae in the LMC. The observed trend is consistent with previous studies of Galactic \citep{shafter2011spitzer,williams2013rapid}.}
    \label{t2td}
\end{figure}

\begin{figure}
    \centering
    \includegraphics[width=1.0\linewidth]{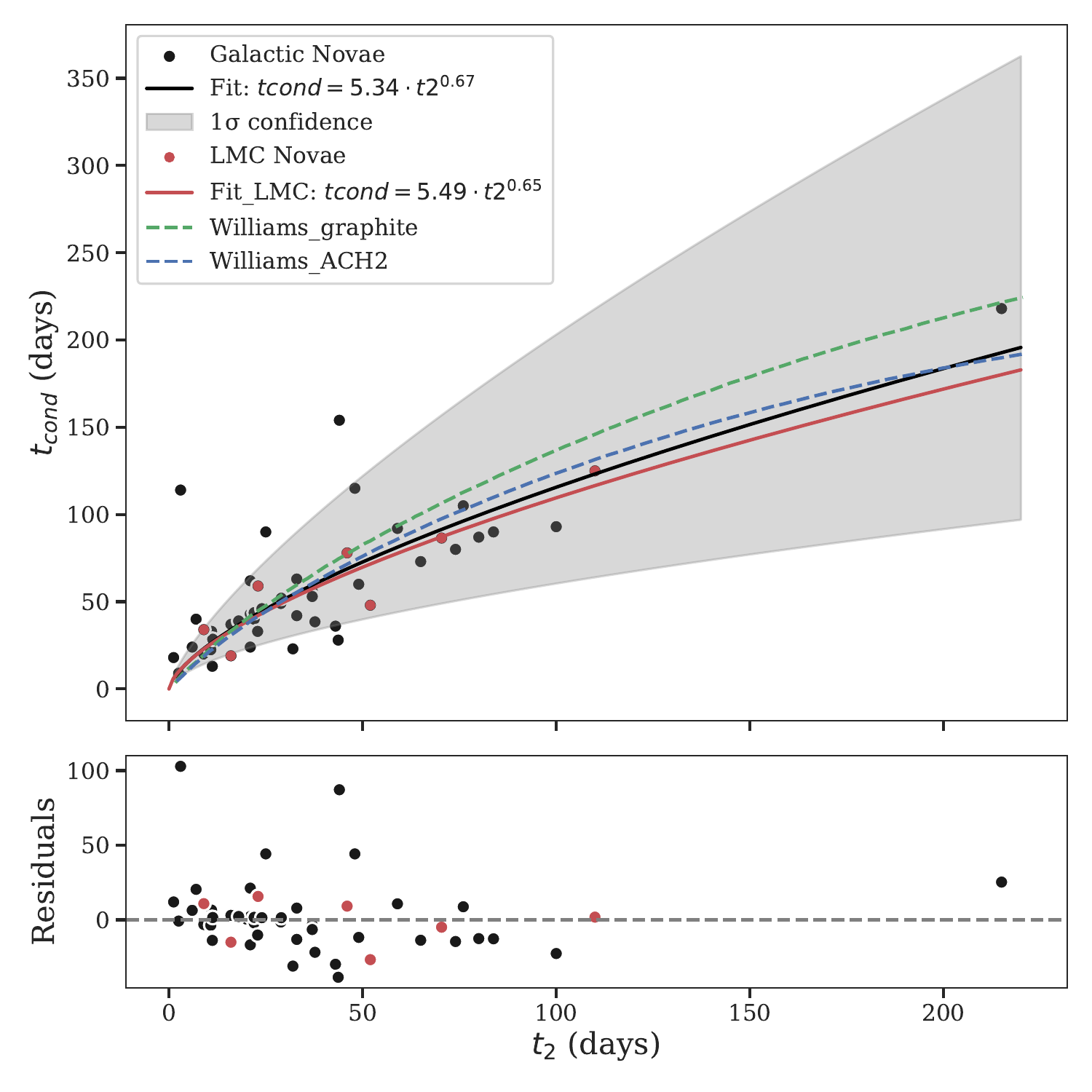}
    \caption{Power-law fit to the correlation between $t_{\mathrm{cond}}$ and $t_2$ for both LMC and Galactic novae. The black solid line represents the best-fit relation, $t_{\mathrm{cond}} = 5.34 \times t_2^{0.67}$, while the shaded region indicates the $1\sigma$ confidence interval. The lower panel displays the residuals of the fit. The red solid line represents the fit for LMC novae only, $t_{\mathrm{cond}} = 5.49 \times t_2^{0.65}$. The dashed green and blue lines show the derived fits from \citet{williams2013rapid} for graphite and ACH2 grains, respectively.}
    \label{powerfit}
\end{figure}

\section{SUMMARY AND CONCLUSIONS}\label{summary}

We have presented observations of nova LMCN 2009-05a, which was discovered in the LMC on HJD 2454956.5 (2009 May 4.9 UT). Optical, and NIR data of this nova have led us to the following conclusions:

\begin{enumerate}

\item LMCN 2009-05a  was a moderately fast nova with t$_2$ = 46 d, and a low mass WD (M$_{WD}$ $\sim$ 0.77 M$_\odot$ ).
We estimated an absolute magnitude M$_V$ = -6.65, which implies that LMCN 2009-05a was a relatively low luminosity
nova with an outburst luminosity of $\sim$ (4.64 ± 0.65)× 10$^{4}$ L$_\odot$.

\item The optical and NIR light curves suggest that a significant amount of dust formed in the nova ejecta between 78 and 155 days after the outburst. The minima of the optical dip were observed on day 108 in the B band and on day 126 in the V, R, and I bands.

\item A total of 34 low-dispersion spectra were presented, covering days 8 to 261 and encompassing the pre-maximum, early decline, and nebular phases. The evolution of P-Cygni profiles indicated slowly moving ejecta. Both permitted and forbidden lines during the early decline phase suggested an inhomogeneous density structure. During the optical minimum, a discrete absorption feature centered at approximately +300 $\kms$ was observed in the H$\alpha$ and [O I] line profiles. The onset of the nebular phase was marked by the emergence of [O III] forbidden lines around day 157.

\item CLOUDY photoionization modeling was employed to perform an abundance analysis of the nova ejecta. A total of 27 emission lines from the day~79 spectrum and 22 lines from the day~236 spectrum were used to obtain the best-fit parameters. The results indicate nitrogen and oxygen abundances significantly enhanced relative to solar values. The temperature of the central source was estimated to be \(1.61 \times 10^4\,\mathrm{K}\) on day~79 and \(2.00 \times 10^5\,\mathrm{K}\) on day~236, with an average luminosity of \(7.38 \times 10^{36}\,\mathrm{erg\,s^{-1}}\).

\item Using the WISE magnitudes, we found that the dust temperature was approximately \(700 \pm 50\,\text{K}\) on day 395. The dust mass is estimated to be \(\sim 2.24 \times 10^{-9}\,M_{\odot}\) and \(\sim 5.05 \times 10^{-9}\,M_{\odot}\), with grain radii of \(0.06 \pm 0.01\,\mu\text{m}\) and \(0.13 \pm 0.04\,\mu\text{m}\) for AC and GR grains, respectively. 

\item We investigated the correlation between the $t_{\text{cond}}$ and \(t_2\) for all dust-forming novae in the LMC, and found that they exhibit behavior consistent with that of Galactic novae.

\end{enumerate}

\begin{table}
\centering
\caption{Dust Formation Timescales of Galactic Novae}
\label{tcond}
\resizebox{\hsize}{!}{%
\begin{tabular}{lccccc}
\hline
\hline
Nova & Year & $t_2$ (days) & $t_{\text{cond}}$ (days) & References \\
\hline
T Aur     & 1892.0  & 80 & 87 & 1 \\[0.25ex]
V476 Cyg  & 1920.6  & 7 & 40 & 1 \\  [0.25ex]
DQ Her    & 1934.9  & 76 & 105 & 1 \\[0.25ex]
V732 Sgr  & 1936.4  & 65 & 73 & 1 \\[0.25ex]
FH Ser    & 1970.1  & 49 & 60 & 1\\[0.25ex]
NQ Vul    & 1976.8  & 21 & 62 & 1\\[0.25ex]
LW Ser    & 1978.2  & 32 & 23 & 1\\[0.25ex]
V1668 Cyg & 1978    & 11 & 33 & 1\\[0.25ex]
V1370 Aql & 1982.1  & 15 & $< 16$ & 1 \\[0.25ex]
PW Vul    & 1984    & 44 & 154 & 1\\[0.25ex]
QU Vul    & 1984    & 20 & 40  & 1\\[0.25ex]
OS And    & 1986.9  & 11 &  22 & 1 \\[0.25ex]
V842 Cen  & 1986.9  & 43 & 36  & 1 \\[0.25ex]
V827 Her  & 1987    & 21 & 43  & 1\\[0.25ex]
QV Vul    & 1987.9  & 37 & 56  & 1\\[0.25ex]
V838 Her  & 1991    & 1.2 & 18 & 1\\[0.25ex]
V992 Sco  & 1992.4  & 100 & 93 & 1 \\[0.25ex]
V1419 Aql & 1993.4  & 25 & 90 & 1 \\[0.25ex]
V705 Cas  & 1993.9  & 33 & 63  & 1 \\[0.25ex]
V445 Pup  & 2000.9  & 215 & 218 & 1 \\[0.25ex]
V2274 Cyg & 2001.5  & 22 & 40 & 1 \\[0.25ex]
V475 Sct  & 2003    & 22 & 43.7 & 2\\[0.25ex]
V2361 Cyg  & 2005    & 6  &  24 & 3, 4\\[0.25ex]
V1663 Aql  & 2005    & 16  &  $< 78$ & 5, 6\\[0.25ex]
V1065 Cen  & 2007    & 11  &  28 & 7\\[0.25ex]
V2615 Oph & 2007    & 33 & 42  & 8\\[0.25ex]
V1280 Sco & 2007.1  & 21 & 24 & 9 \\[0.25ex]
V5579 Sgr & 2008.3  & 9  & 20 & 10 \\[0.25ex]
QY Mus    & 2008.7  & 48 & 115 & 11, 12 \\[0.25ex]
V496 Sct  & 2009.8  & 59 & 92 & 13\\[0.25ex]
V1368 Cen & 2012    & 16 & 36 & 2\\[0.25ex]
V1428 Cen & 2012    & 10.8 & 22.5 & 2\\[0.25ex]
V1324 Sco & 2012.5  & 24 & 46 & 2,14\\[0.25ex]
V5592 Sgr & 2012.6  & $< 16.8$ & 30 & 2 \\[0.25ex]
V2676 Oph & 2012.3  & 83.8 & 90 & 2, 15 \\[0.25ex]
V809 Cep  & 2013.2  & 16 & 37 & 16, 17 \\[0.25ex]
V1533 Sco & 2013    & 11 & 14 & 18\\[0.25ex]
V339 Del  & 2013    & 11.3 & 28.5 & 2\\[0.25ex]
V745 Sco  & 2014    & 2.5  &  9 & 2, 19\\[0.25ex]
V1369 Cen & 2014.0  & 37.7 & 38.5 & 2 \\[0.25ex]
V5668 Sgr & 2015.3  & 74 & 80 & 2, 20 \\[0.25ex]
V1831 Aql & 2015.8  & 16.4  & 19.2  & 21 \\[0.25ex]
V1655 Sco & 2016.5  & 28.9 & 49 & 22, 12,\\[0.25ex]
V3662 Oph & 2017    & 37   &  53 & 2, 23\\[0.25ex]
V1661 Sco & 2018    & 11.2  &  13 & 2\\[0.25ex]
V906 Car  & 2018    & 43.7  &  28 & 2\\[0.25ex]
V357 Mus  & 2018.1  & 22.9 & 33 & 2 \\[0.25ex]
V2891 Cyg  & 2019    & 100–150  &  273 & 24\\[0.25ex]
V1391 Cas & 2020.7  & uncertain due to flares & 141 & 25 \\[0.25ex]
V1112 Per & 2021.0  & 18 & 39 & 26, 27 \\[0.25ex]
V6594 Sgr & 2021.3  & 26 & 52 & 12, 28  \\[0.25ex]
V606 Vul  & 2021.6  & 3  & 114 & 29\\[0.25ex]
\hline
\end{tabular}
}
\vspace{0.5em}
\noindent \textbf{References.} (1) \cite{shafter2011spitzer}; (2) \cite{2025arXiv250104098C}; (3) \cite{2007ApJ...662..552H};(4) \cite{2025RNAAS...9...14R}; (5) \cite{ness2007swift}; (6) \cite{2024RNAAS...8..317R}; (7) \cite{2010AJHelton}; (8) \cite{2009MNRAS.398..375D}; (9) \cite{2008MNRAS.391.1874D}; (10) \cite{raj2024dustyaftermathrapidnova}; (11) \cite{2018A&A...609A.120F}; (12) This paper; (13) \cite{2012MNRAS.425.2576R}; (14) \cite{2018ApJ...852..108F}; (15) \cite{raj2017v2676}; (16) \cite{2014MNRAS.440.3402M}; (17) \cite{2022MNRAS.515.3028B}; (18) \cite{2021RNAAS...5..120R}; (19) \cite{2023ApJ...954L..16B}; (20) \cite{2022MNRAS.511.1591T}; (21) \cite{2018MNRAS.473.1895B}; (22) \cite{kawash2021classical}; (23) \cite{2017ATel10492....1J}; (24) \cite{2022MNRAS.510.4265K}; (25) \cite{2020ATel14272....1B}; (26) \cite{2021RNAAS...5..273R}; (27) \cite{2021ATel14338....1B}; (28) \cite{2021ATel14637....1M}; (29) \cite{2023arXiv231104903S}

\end{table}

\section*{Acknowledgements}
The authors would like to thank the referee for critically
reading the manuscript and providing valuable suggestions
to improve it. We thank Dr. Shrish Raj, NTU, Singapore  for valuable discussion.
 We acknowledge with thanks the variable star observations from the AAVSO International Database contributed by observers worldwide and used in this research. We also acknowledge the use of SMARTS data.

\bibliography{sample631}{}
\bibliographystyle{aasjournal}



\end{document}